\documentclass[aps,prd,preprintnumbers,10pt,twocolumn,superscriptaddress,showpacs,longbibliography]{revtex4-1}

\pdfoutput=1
\usepackage{graphicx}
\usepackage{amsfonts,amsmath,amssymb,bm,bbm}
\usepackage{enumitem}
\usepackage{url}  
\usepackage{hyperref}
\usepackage{subcaption}
\usepackage{ragged2e}

\usepackage{comment}
\usepackage{mathtools}  
\usepackage{cleveref}
\usepackage{color}
\usepackage{xcolor}
\usepackage[dvipsnames]{xcolor}
\usepackage[normalem]{ulem}
\usepackage{physics}

\usepackage{multirow}
\usepackage{booktabs}

\definecolor{darkgreen}{HTML}{008000}

\makeatletter
\newcommand{\Rmnum}[1]{\expandafter\@slowromancap\romannumeral #1@}

\makeatother

\hypersetup{
    pdfnewwindow=true,      
    colorlinks=true,       
    linkcolor=blue,          
    citecolor=blue,        
    filecolor=blue,      
    urlcolor=blue        
}

\def\bi{\begin{itemize}[noitemsep,leftmargin=2em]
\setlength\itemsep{0.1em}
        }
\def\ei{\end{itemize}}

\newcommand{\orcid}[1]{\begingroup
  \hypersetup{hidelinks}\href{https://orcid.org/#1}{\includegraphics[width=10pt]{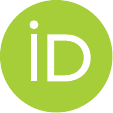}} \endgroup}

\newcommand{\be}{\begin{equation}\begin{aligned}}
\newcommand{\ee}{\end{aligned}\end{equation}}

\begin{document}

\title{How Bright in Gravitational Waves are Millisecond Pulsars for the Galactic Center GeV Gamma-Ray Excess? A Systematic Study and Implications for Dark Matter}

\author{Ming-Yu Lei\ \ }
\affiliation{Key Laboratory of Dark Matter and Space Astronomy, Purple Mountain Observatory, Chinese Academy of Sciences, Nanjing 210023, China}
\affiliation{School of Astronomy and Space Science, University of Science and Technology of China, Hefei 230026, China}

\author{Bei Zhou \orcid{0000-0003-1600-8835}\,}
\email{beizhou@fnal.gov}
\affiliation{Theory Division, Fermi National Accelerator Laboratory, Batavia, Illinois 60510, USA}
\affiliation{Kavli Institute for Cosmological Physics, University of Chicago, Chicago, Illinois 60637, USA}

\author{Xiaoyuan Huang \orcid{0000-0002-2750-3383}\,}
\email{xyhuang@pmo.ac.cn}
\affiliation{Key Laboratory of Dark Matter and Space Astronomy, Purple Mountain Observatory, Chinese Academy of Sciences, Nanjing 210023, China}
\affiliation{School of Astronomy and Space Science, University of Science and Technology of China, Hefei 230026, China}

\preprint{FERMILAB-PUB-25-0822-T}

\date{\today}

\begin{abstract}
The existence of dark matter (DM) is supported by various macroscopic observations, but its microscopic nature remains elusive. 
The Galactic Center GeV gamma-ray excess (GCE) has been a leading candidate signal for particle DM annihilation. 
However, an unresolved population of millisecond pulsars (MSPs) in the bulge provides the alternative explanation for the excess.
Identifying these MSPs in electromagnetic bands is difficult due to source confusion, pulse broadening, and extinction. 
Gravitational waves (GWs) provide a complementary probe: a steadily rotating, nonaxisymmetric MSP emits a nearly monochromatic GW signal in the sensitive band of ground-based detectors, with amplitude set by its ellipticity.
In this work, we systematically investigate the GW emission from the MSP population proposed to explain the GCE and its detectability with current and future detectors.
We consider three major scenarios for the origin of ellipticity and model the population properties of these MSPs.
We also consider both isolated MSPs and MSPs in binary systems, as well as Doppler effects in the detection. 
We find that while the signal is generally below the reach of current interferometers, next-generation detectors such as the Einstein Telescope and Cosmic Explorer may detect a fraction of those MSPs, offering a novel test of the MSP interpretation of the GCE. 
Future directed searches toward the Galactic Center may either uncover this long-sought MSP population or place stringent constraints on bulge MSP properties. Continued improvements in detector sensitivity will provide increasingly stringent tests of the MSP interpretation, with growing power to distinguish between the astrophysical and DM interpretations of the GCE.
\end{abstract}

\maketitle


\section{Introduction}
\label{sec_intro}

The existence of dark matter (DM) is supported by a broad set of macroscopic observations, including galaxy rotation curves~\cite{Rubin:1980zd,Bosma:1981zz,Corbelli:1999af}, lensing in clusters (including the Bullet Cluster)~\cite{Clowe_2004,Clowe:2006eq}, large-scale structure~\cite{SDSS:2005xqv, Springel:2006vs}, and precision measurements of the cosmic microwave background~\cite{WMAP:2012nax,Planck:2018vyg}.
Despite this breadth of macroscopic evidence and an increasingly precise determination of the cosmic DM abundance, the microscopic nature of DM remains unknown: no experiment has yet identified the particle(s) that constitute DM or measured their interactions beyond gravity~\cite{Jungman:1995df,Bertone:2016nfn,Cirelli:2024ssz}.
Worldwide experiments conducting direct, indirect, and collider searches continue to narrow viable parameter space across many candidate frameworks (e.g., weakly interacting massive particles [WIMPs], axions, sterile neutrinos), but no conclusive signal has been established~\cite{Bulbul:2014sua,Fermi-LAT:2015att,ATLAS:2017nga,ADMX:2018gho,Billard:2021uyg,PandaX-4T:2021bab,An:2022hhb,Guo:2022rqq,LZ:2022lsv,XENON:2023cxc, LHAASO:2024upb}.

The Galactic Center GeV gamma-ray excess (GCE) has been a leading candidate signal for particle DM.
It is characterized by a bump spectrum centered around 2 GeV with a several-GeV width and a roughly spherically symmetric morphology extending several degrees around the Galactic Center (GC). 
It was first identified and interpreted as a possible signature of DM annihilation in Ref.~\cite{Goodenough:2009gk}.
While the presence of this excess has been consistently reported across multiple independent analyses, substantial debate remains regarding its morphology and spectral properties, owing to significant uncertainties in modeling the Galactic diffuse foreground~\cite{Macias:2013vya, Abazajian:2014fta, Zhou:2014lva, Calore:2014xka, Huang:2015rlu, Fermi-LAT:2017opo, Zhong:2019ycb, Cholis:2021rpp, Zhong:2024vyi, Song:2024iup}.
This sensitivity to foreground modeling affects the relative viability of the two competing interpretations.
A variety of particle DM models have been invoked to explain the GCE~\cite{Hooper:2010mq, Daylan:2014rsa, Calore:2014nla, Agrawal:2014oha, Izaguirre:2014vva, Leane:2019xiy, Murgia:2020dzu}, most prominently WIMPs with masses of a few tens of GeV annihilating predominantly into $b\bar b$, which hadronize and decay to gamma rays.

However, there is also a well-motivated astrophysical interpretation: an unresolved population of MSPs in the inner Galaxy can reproduce the observed GCE spectrum and morphology at current \textit{Fermi}-LAT resolution~\cite{Abazajian:2010zy,Yuan:2014rca,Petrovic:2014xra,Lee:2015fea,Bartels:2015aea,Macias:2016nev,Bartels:2017vsx,Gautam:2021wqn}\footnote{We note that some earlier studies questioned the MSP interpretation, arguing that the spectra of known MSPs are too soft at sub-GeV energies and that population estimates inferred from low-mass X-ray binary counts suggest unresolved MSPs could contribute only a small fraction of the GCE intensity~\cite{Hooper:2013nhl,Cholis:2014lta}. However, subsequent analyses have largely addressed these concerns. The sub-GeV spectrum of the GCE is now known to be heavily affected by systematic uncertainties in Galactic diffuse emission modeling~\cite{Fermi-LAT:2017opo}. Furthermore, recent binary population synthesis models have demonstrated that an MSP population formed via the accretion-induced collapse of oxygen-neon white dwarfs can naturally reproduce the GCE's intensity, spectrum, and morphology, while strictly obeying LMXB constraints~\cite{Gautam:2021wqn}.}.
These MSPs are mostly in the Galactic bulge, so hereafter we refer to them as \textbf{``bulge MSPs''} for simplicity.
MSPs are rapidly rotating neutron stars (NSs) whose rotational kinetic energy can power radio, gamma rays, and other electromagnetic emissions.
Young pulsars spin down on Myr timescales; subsequent accretion can recycle ``dead'' pulsars to millisecond periods~\cite{1982Natur.300..728A}.
These recycled MSPs typically have lower surface magnetic fields, $B_{\rm surf}\sim10^8$--$10^9$~G, and their long lifetimes of order Gyr make the existence of a substantial bulge population astrophysically plausible~\cite{BHATTACHARYA19911,Lorimer:2008se,Gnedin:2013cda, Arca-Sedda:2014ila, Brandt:2015ula, Macquart:2015jfa}.

Identifying these bulge MSPs would provide a direct way to distinguish the DM and MSP interpretations of the GCE. However, resolving these sources across electromagnetic bands remains challenging, and only one candidate, PSR J1744-2946~\cite{Lower:2024sdi}, has been tentatively identified so far.
In gamma rays, most bulge MSPs are too faint to resolve individually, and the $\sim$degree-scale point-spread function at GeV energies leads to severe source confusion~\cite{Fermi-LAT:2021wbg}; only a few candidates have been discussed, and none are confirmed~\cite{Ploeg_2020,Holst:2024fvb}.
In radio, the detectability of bulge MSPs is limited by both strong multipath scattering, which broadens pulsar signals and makes timing searches difficult, and bright foreground emission in the Galactic plane~\cite{Cordes:2002wz,2013CQGra..30v4003S,Calore:2015bsx,Jing:2025fxl}.
In UV/optical/near-IR, extinction toward the GC is substantial and isolated MSPs are intrinsically faint. 
Analyses of static x-ray properties (e.g., flux and spectra), combined with multiwavelength data, can identify promising MSP candidates in the Galactic bulge~\citep{Berteaud:2020zef, Berteaud:2024rgv}. However, such methods do not yield timing information; confirmation still relies on targeted pulse searches to detect the characteristic pulsations.

Gravitational waves (GWs) provide a complementary probe that is insensitive to dust, scattering, and the source-confusion problems in the crowded stellar fields around the GC. 
A steadily rotating, nonaxisymmetric neutron star emits a continuous, nearly monochromatic GW signal at $f_{\rm GW}\simeq 2 f_{\rm rot}$, typically in the $\simeq 200-2000$~Hz range for MSPs, which coincides with the most sensitive band of ground-based interferometers~\footnote{
In addition, the mHz band of the Laser Interferometer Space Antenna could also probe those MSPs in the binary systems; see Ref.~\cite{Korol:2025fkk} for a recent study.
}, 
including LIGO~\cite{LIGOScientific:2014pky}, Virgo~\cite{VIRGO:2014yos}, KAGRA~\cite{KAGRA:2018plz}, and next-generation Cosmic Explorer (CE)~\cite{Evans:2023euw} and Einstein Telescope (ET)~\cite{Punturo:2010zz}.
Such GW emission is persistent and nearly monochromatic, in stark contrast to the transient broadband signals from compact-binary coalescences.
Although it is very difficult to resolve the large number of MSPs in the crowded bulge region using gamma rays and other electromagnetic wavelengths, GWs can potentially resolve individual MSPs in the frequency domain, thanks to a resolution of $\Delta f \sim 1/T_{\rm obs}$.

In this paper, we present a systematic analysis that incorporates a physically motivated signal prediction and a detailed treatment of the detector response, building upon previous work~\cite{Calore:2018sbp, Miller:2023qph, Bartel:2024jjj}.
We outline the main ingredients of our analysis in Table~\ref{summary_table}. This framework allows us to investigate the GW emission expected from the MSP population responsible for the GCE and assess its detectability with present and future ground-based GW interferometers~\footnote{
As a complementary avenue to ground‑based interferometers, a recent proposal suggests probing kHz‑band backgrounds using a laser-polarization ``Gravitational Photon Polarization Twist'' in space~\cite{Capdevilla:2025ivz}.
With optimistic baselines ($\sim 10^6$ km) and millisecond pulses, its reach could overlap the collective strains expected from a bulge MSP population~\cite{Capdevilla:2025ivz}.}.
The GW emission arises from a finite ellipticity, and we consider major physically motivated scenarios that can generate such ellipticities: 
(1) internal magnetic-field-induced deformation, 
(2) long-lived crustal mountains from breaking strain, and 
(3) a general parametrization in terms of a fraction of the pulsar spin-down energy. 
Importantly, the third scenario also naturally yields a \textit{theoretical maximum}, assuming all the spin-down power is converted to GW emission.
We model the bulge MSPs by combining a fixed boxy bulge morphology, constrained by the GCE, with two distinct population frameworks, an Empirical Disk-Proxy Model and an Evolutionary Model, which respectively derive the necessary intrinsic MSP parameters and total population size empirically and self-consistently.
For detectability, we outline two search strategies---coherent and incoherent---that explicitly account for Doppler shifts from Earth’s motion and binary orbits.
Our results demonstrate that, under the assumptions considered here, the bulge MSPs can produce GW signals approaching the sensitivity of next-generation ground-based interferometers, such as CE and the ET~\cite{Srivastava:2022slt,Hild:2008ng}. For the theoretical maximum, a fraction of MSPs can reach the LIGO-Virgo-KAGRA (LVK) O4 sensitivity~\cite{Capote:2024rmo,LIGOScientific:2025kei}.

This paper is organized as follows: in Sec.~\ref{sec_gw}, we discuss the continuous GW emission from MSPs and define the key quantities used throughout. 
In Sec.~\ref{sec_ellipticity}, we present three major scenarios for the origin of MSPs’ ellipticity. 
In Sec.~\ref{sec_pop}, we describe the bulge MSP population models, including intrinsic parameters and spatial morphologies. We also discuss the total number of bulge MSPs in light of the GCE. 
In Sec.~\ref{sec_detect}, we discuss the detection of MSPs' GW signals, including the Doppler shifts in frequency, and outline two search strategies: coherent and incoherent. 
In Sec.~\ref{sec_results}, we present our results, including their GW strain amplitude and detectability in current and next-generation GW detectors. We conclude in Sec.~\ref{sec_concl}.

\section{GW emissions from MSPs}
\label{sec_gw}

\begin{figure*}[tbp]
\makeatletter\def\@captype{table}\makeatother
\centering
\setlength{\tabcolsep}{4pt}
\renewcommand{\arraystretch}{1.4}
\begin{tabular}{llll}
\toprule
\hline\hline
\textbf{Ingredients} & \textbf{Cases}  & \textbf{Brief summaries} & \textbf{Sections} \\
\hline
\multirow{3}{*}{\shortstack[l]{Ellipticity \\ origins}} 
& Magnetic deformation                  & $\epsilon \propto B_{\rm{internal}} \propto B_{\rm{surface}}$                    & \ref{sec_ellipticity_MF} \\
& Crustal mountains                     & $\epsilon = 10^{-9}$                                                             & \ref{sec_ellipticity_crustal} \\
& Spin-down energy fraction             & $\epsilon \propto \sqrt{\eta}$, from $\dot{E}_{\rm GW} = \eta \dot{E}_{\rm rot}$ & \ref{sec_ellipticity_Eloss} \\
\hline
\multirow{2}{*}{\shortstack[l]{Population \\ models}} 
& Pop.~\Rmnum{1}                        & $N_{\rm MSP} \simeq 2.9 \times 10^5$ (ATNF catalog~\cite{ATNF_catalog} \& Ref.~\cite{List:2025qbx})   & \multirow{2}{*}{\ref{sec_pop_paras}, \ref{sec_pop_NMSP}} \\
& Pop.~\Rmnum{2}                        & $N_{\rm MSP} \simeq 1.6 \times 10^4$ (Ploeg et al.~\cite{Ploeg_2020})   &  \\
\hline
\multirow{2}{*}{Morphology} 
& Boxy bulge                            &      \multirow{2}{*}{GW forecasts nearly morphology-independent}        & \ref{sec_pop_morphology} \\
& Spherical bulge                       &        & Appendix~\ref{sec_app_morpho_dep} \\
\hline
\multirow{2}{*}{\shortstack[l]{Detection \\ strategies}} 
& Coherent                              & Long coherent integration; Doppler modulation correctable                           & \ref{sec_detect_simplest}, \ref{sec_detect_ourwork} \\
& Incoherent                            & Stacking short segment; Doppler modulation uncorrectable     & \ref{sec_detect_doppler}, \ref{sec_detect_ourwork} \\
\hline\hline
\bottomrule
\end{tabular}
\caption[OptionalShort]{\justifying 
Summary of the ingredients of this work's analysis.
}
\label{summary_table}
\end{figure*}

To estimate the GW signals from MSPs, we adopt the standard quadrupolar emission formula, assuming a steadily rotating, nonaxisymmetric rigid neutron star, with a constant ellipticity, radiating in the far field under the linearized theory of general relativity~\cite{Maggiore:2007ulw}. These are well motivated and widely used assumptions. Accordingly, 
\begin{equation}
\begin{aligned}
h_{+} &= h_0 \frac{1+\cos^2 \iota}{2} \cos (2\pi f_{\mathrm{GW}} t) \,, \\
h_{\times} &= h_0 \cos \iota \sin (2\pi f_{\mathrm{GW}} t) \,,
\label{eq_h}
\end{aligned}
\end{equation}
where
\begin{equation}
h_0 = \frac{4 \pi^2 G}{c^4} \frac{I_{zz} f_{\mathrm{GW}}^2}{r} \epsilon \,.
\label{eq_h0}
\end{equation}
Here, $h_+$ and $h_\times$ are the two GW polarization amplitudes,
$\iota$ is the inclination angle between the MSP’s rotation axis and the line of sight,
$f_{\mathrm{GW}}$ is the GW frequency (twice the pulsar’s rotational frequency, $f_{\rm rot}$),
$G$ is Newton’s gravitational constant,
$c$ is the speed of light,
$I_{zz}$ is the moment of inertia about the rotation axis,
$r$ is the distance to the pulsar,
and $\epsilon$ is the equatorial ellipticity, with 
\be
\epsilon \equiv \frac{I_{xx}-I_{yy}}{I_{zz}} \,,
\ee
i.e., the fractional difference between the two moments of inertia orthogonal to the rotation axis.

NSs typically have a mass of $m_{\rm NS} \simeq 1.4 M_{\odot}$ and a radius of $R_{\rm NS} \simeq$ 12~km, which gives $I_{zz} \simeq(2 / 5) m_{\rm NS} R_{\rm NS}^2 \simeq 1.6 \times 10^{38} \, \mathrm{kg}\,\mathrm{m}^2$. 
As a result, 
\be
h_0 \simeq 2.1 \times 10^{-28}
\left(\frac{\epsilon}{10^{-9}}\right)
\left(\frac{m_{\rm NS}}{1.4\, M_\odot} \right) 
\left(\frac{R_{\rm NS}}{12 \, {\rm km}} \right)^2  \\ \times
\left(\frac{8.178 \, \mathrm{kpc}}{r}\right)
\left(\frac{f_{\mathrm{GW}}}{1\, \mathrm{kHz}}\right)^2.
\ee

\section{Origins of MSPs Ellipticity}
\label{sec_ellipticity}

A key parameter governing the emission of continuous GWs from MSPs is the stellar ellipticity, $\epsilon$. This quantity describes the deviation of the NS from perfect axisymmetry. 
In this work, we explore three distinct scenarios, as listed in Table~\ref{summary_table}:
(1) ellipticity generated by stresses from strong internal magnetic fields within the NS; 
(2) ellipticity arising from nonaxisymmetric crustal deformations (“crustal mountains”) that become permanently frozen in as the NS cools and the solid crust forms;
and (3) a model-independent parametrization in which GW emission accounts for a fraction $\eta$ of the total spin-down energy. 
The first two scenarios are based on specific physical mechanisms and may contribute simultaneously to the ellipticity~\cite{Pagliaro:2023bvi}.
Importantly, the third scenario provides a \textit{theoretical maximum} when $\eta = 100\%$, i.e., GW emission accounts for all the spin-down power.

Previous work by Woan et al.~\cite{Woan:2018tey}, using MSP data available up to 2018, suggested that MSPs may sustain a minimum ellipticity at the level of $\sim 10^{-9}$. 
However, this conclusion is derived from the existing MSP observations and may be influenced by the observational selection effect. The confirmation of additional MSPs exhibiting short rotation periods and extremely low period derivatives, beyond the two already documented in Ref.~\cite{Woan:2018tey}, would challenge the current hypothesis.
Thus, we do not apply a minimum ellipticity in our calculations.

This section focuses on the physical origins of ellipticity in individual MSPs. The application of these models to the MSP population in the bulge, including the resulting ellipticity distributions, is discussed in Sec.~\ref{sec_pop}.

\subsection{Magnetic-Field-Induced Deformation}
\label{sec_ellipticity_MF}

A strong internal magnetic field within an NS can induce a nonaxisymmetric deformation. If the axis of the internal magnetic field, $B_{\rm int}$, is misaligned with the star's rotation axis, the resulting magnetic pressure is anisotropic, distorting the star from a perfect spheroid~\cite{Chandrasekhar:1953zz, Maggiore:2018sht}. Theoretical estimates also suggest that the protons in the NS core form a type II superconductor~\cite{MIGDAL1959655,1969Natur.224..872B}. In such a state, the internal magnetic field is confined to quantized flux tubes. The tension from these flux tubes creates an anisotropic stress that deforms the star. 
This magnetic-field-induced deformation results in an ellipticity that scales linearly with the internal magnetic field strength,  $B_{\rm{internal}}$~\cite{Cutler:2002nw,Akgun:2007ph, Lander_2011, Lander:2013oea, LIGOScientific:2020gml}, as follows: 
\begin{equation} 
\epsilon \approx 10^{-8} \left(\frac{B_{\rm{internal}}}{10^{12} ~ \rm{G}}\right). 
\label{eq_ellipticity_from_Bfield}
\end{equation}

While the internal magnetic field, $B_{\rm{internal}}$, cannot be directly observed, it is commonly inferred from the measured external surface dipolar magnetic field $B_{\rm{surface}}$. The ratio of $B_{\rm{internal}}$ to $B_{\rm{surface}}$ is expected to be $\sim 10^2$ to $10^4$ for MSPs~\cite{Mastrano:2011aa, Ciolfi:2013dta, Haskell:2015psa}.  Consistent with previous studies~\cite{Miller:2023qph, Bartel:2024jjj}, we adopt a conservative ratio for this work:
\begin{equation}
B_{\rm{internal}} = 150 B_{\rm{surface}}.
\label{eq_ratio_Bfield}
\end{equation}

The surface magnetic field can be inferred from the observed period ($P$) and its derivative ($\dot{P}$). We move beyond the simple vacuum dipole approximation, which is insufficient for the plasma-filled environments of pulsars, and adopt a relation derived from the braking torque in a force-free magnetosphere model ~\cite{Spitkovsky:2006np}. The surface magnetic field is thus given by
\begin{equation}
    B_{\rm surf}^2 = \frac{3 c^3 I_{zz} P \dot{P}}{5 \pi^2 R_{\rm NS}^6},
\label{eq_Bsurf}
\end{equation}
where $I_{zz}$ is the moment of inertia and we assume a canonical NS radius $R_{\rm NS} = 12$~km. Combining these relations, we obtain a scaling for the ellipticity as a function of the observable MSP parameters, as follows:
\be
\epsilon \simeq 
2.7 \times 10^{-10}\left(\frac{f_{\mathrm{rot}}}{300 \mathrm{~Hz}}\right)^{-\frac{3}{2}}
\left(\frac{\left|\dot{f}_{\mathrm{rot}}\right|}{10^{-15} \mathrm{~Hz} \, \mathrm{s}^{-1}}\right)^{\frac{1}{2}} \\ \times
\left( \frac{m_{\rm NS}}{1.4 M_\odot} \right)^{\frac{1}{2}}
\left( \frac{12\, {\rm km}}{R_{\rm NS}} \right)^2 \,,
\label{eq_ellipticity_from_Bfield_scaling}
\ee
where $f_{\rm rot}$ and $\dot{f}_{\rm rot}$ are the rotational frequency and its first time derivative, and $m_{\rm NS}$ is the NS mass.

\subsection{Crustal Mountains from Breaking Strain}
\label{sec_ellipticity_crustal}

During the recycling of an MSP, accretion can produce large-scale interior temperature asymmetries misaligned from the spin axis. These gradients drive temperature-sensitive electron captures in the deep crust that provide a mass quadrupole, often referred to as a mountain. This accretion-driven structure offers a natural mechanism for GW emission, limited primarily by the crust's breaking strain~\cite{Bildsten:1998ey, Ushomirsky:2000ax, Owen:2005fn, Andersson_2010, Johnson-McDaniel:2012wbj}.

In the extremely ideal scenario, if the whole crust reaches the breaking strain, the maximum ellipticity is reached at 
\be
\epsilon^{\rm ideal} < 4 \times 10^{-6} \left( \frac{u_{\text{break}}}{0.1} \right),
\ee
where $u_{\rm break}$ is the crustal breaking strain (distinct from the ``strain'' in GW detection), and is $\sim0.1$ determined from molecular-dynamics simulations of Coulomb solids representative of the NS crust~\cite{Horowitz:2009ya}. 

In practice, however, only localized regions of the crust reach stresses close to the breaking strain.
As a result, the typical initial ellipticity is expected to be significantly smaller than the theoretical maximum, at the level of 
\be
\epsilon^{\rm initial} \sim 10^{-8} \text{ to } 10^{-7} \,.
\ee

Over longer timescales, plastic flow and viscoelastic creep gradually relax these deformations~\cite{Chugunov:2010ac}, reducing the ellipticity to a long-term residual value of ~\cite{Morales:2024hwl}
\be
\epsilon \sim 10^{-9} \,,
\label{eq_ellipticity_from_CM}
\ee
which can last for gigayears and is the realistic target for GW emission from MSPs.
This expectation is consistent with recent results indicating that a slightly anisotropic neutron
star crust, undergoing modest changes in rotation rate, can support an ellipticity of a few\,$\times$\,$10^{-9}$~\cite{Morales:2024zka}, comparable to the observational upper bounds inferred for some nearby and rapidly spinning pulsars~\cite{KAGRA:2022osp}.

Therefore, we adopt a constant typical ellipticity of $\epsilon = 10^{-9}$. This is independent of MSPs' parameters. 

\subsection{Model-Independent Spin-Down Energy Fraction}
\label{sec_ellipticity_Eloss}

We also consider a more model-independent way to parametrize the ellipticity by simply assuming that a fraction, $\eta$, of the spin-down energy loss goes into GW emission.

The ellipticity in this scenario can be solved as follows:
the rotational energy loss due to GW emission is given by~\cite{Maggiore:2007ulw}
\begin{equation}
\eta \frac{\dd E_{\mathrm{rot}}}{\dd t}=-\frac{32 G}{5 c^5} \epsilon^2 I_{zz}^2 \omega_{\mathrm{rot}}^6 \,,
\end{equation}
where $\omega_{\rm rot} \equiv 2 \pi f_{\rm rot} $. The rotational energy of an NS about its principal axis $x_{zz}$ is expressed as $E_{\mathrm{rot}} = (1/2) I_{zz} \omega^2_{\rm rot}$. 
This leads to
\begin{equation}
\epsilon = \sqrt{ \frac{5 c^5}{2 G} \frac{|\dot{f}_{\rm rot}|}{I_{zz} f_{\rm rot}^5}  } \cdot \frac{1}{16 \pi^2} \cdot \sqrt{\eta}.
\label{eq_epsilon_from_eta}
\end{equation}
Here, the ${f}_{\rm rot}$ and $\dot{f}_{\rm rot}$ denote the pulsar's rotational frequency and its time derivative, respectively.

We adopt $\eta = 1\%$.
Plugging in the other typical numbers, we obtain the scaling of $\epsilon$ with MSPs' parameters, as follows:
\be
\epsilon \simeq 
3.1 \times 10^{-10}
\left(\frac{\eta}{0.01}\right)^{1/2}
&  \left(\frac{f_{\mathrm{rot}}}{300 \, \mathrm{Hz}}\right)^{-5/2}
\left(\frac{\left|\dot{f}_{\mathrm{rot}}\right|}{10^{-15} \, \mathrm{Hz/s}}\right)^{1/2} \\ \times
& \left(\frac{1.4 M_\odot}{m_{\rm NS}} \right)^{1/2}
\left(\frac{12 \, {\rm km}}{R_{\rm NS}} \right).
\label{eq_epsilon_scaling_Eloss}
\ee
We emphasize that this choice serves as an illustrative reference rather than a prediction of a specific physical mechanism. According to recent continuous GW searches for nearby MSPs using LVK observations~\cite{LIGOScientific:2020gml,LIGOScientific:2021hvc,LIGOScientific:2025kei}, the most stringent constraint on $\eta$ is at the $\sim 10\%$ level, which is higher than our choice.

Importantly, a {\bf theoretical maximum} on $\epsilon$ can be naturally set in this scenario from the {\bf spin-down limit}, i.e., assuming all ($\eta = 100\%$) the spin-down energy is taken away by GW emission, as follows:
\be
\epsilon^{\max} 
\simeq 3.1 \times 10^{-9} 
\left(\frac{f_{\mathrm{rot}}}{300 \mathrm{~Hz}}\right)^{-5 / 2}
\left(\frac{\left|\dot{f}_{\mathrm{rot}}\right|}{10^{-15} \mathrm{~Hz} / \mathrm{s}}\right)^{1/2} \\ \times
\left(\frac{1.4 M_\odot}{m_{\rm NS}} \right)^{1/2}
\left(\frac{12 \, {\rm km}}{R_{\rm NS}} \right)
\label{eq_epsilon_scaling_Eloss_maximum}
.
\ee

\section{Population Properties of Bulge MSPs}
\label{sec_pop}

In this section, we discuss the population properties of the bulge MSPs that determine their GW detectability, as outlined in Table~\ref{summary_table}. In Sec.~\ref{sec_pop_paras}, we discuss the distributions of the intrinsic parameters of MSPs in the bulge, including magnetic field strength ($B$), rotational frequency ($f_{\rm rot}$) and its time derivative ($\dot{f}_{\rm rot}$), which are directly related to the period ($P$) and its derivative ($\dot{P}$). These parameters, in conjunction with the discussion in Sec.~\ref{sec_ellipticity}, determine the ellipticity distributions of bulge MSPs, as shown in Sec.~\ref{sec_pop_ellipticity}. In Sec.~\ref{sec_pop_morphology}, we discuss the morphological distribution, as the distances to the MSPs affect their GW detectability. Finally, in Sec.~\ref{sec_pop_NMSP}, we estimate the total number of bulge MSPs, $N_{\rm MSP}$, as constrained by their gamma-ray luminosity and the observed GCE luminosity.

\subsection{Models for MSP Population Parameters}
\label{sec_pop_paras}

A statistically significant population of MSPs has yet to be individually resolved in the bulge. Therefore, inferring the properties of this putative population requires model-based approaches, for which we explore two primary, physically distinct frameworks. The first is the \textbf{Empirical Disk-Proxy Model (Pop.~\Rmnum{1})}, which assumes the bulge MSPs share the same present-day properties as the well-characterized population in the Galactic disk, with fiducial distributions derived from the Australian Telescope National Facility (ATNF) pulsar database~\cite{ATNF_catalog}. The second framework is the \textbf{Evolutionary Model (Pop.~\Rmnum{2})}, which is a theoretical approach informed by the work of Ploeg et al.~\cite{Ploeg_2020}. It interprets the GCE as originating from an older MSP population native to the bulge, allowing for a self-consistent derivation of its evolved properties (e.g., longer periods, lower luminosities) from universal birth parameters and the region's specific star formation history~\cite{Crocker:2016zzt}. 

Pop.~\Rmnum{1} provides a realistic, empirically derived set of MSP parameters. While Pop.~\Rmnum{2} is more physically motivated and provides a direct fit to the observed GCE, its predicted number of MSPs, discussed in Sec.~\ref{sec_pop_NMSP}, exhibit mild tension with a recent machine learning analysis~\cite{List:2025qbx}, which builds upon earlier studies~\cite{List:2020mzd,List:2021aer} and favors a larger MSP population of $N_{\rm MSP} > 10^5$. We therefore include both population models for comparison.

We first construct Pop.~\Rmnum{1} by deriving distributions for the rotational frequency ($f_{\rm rot}$) and its derivative ($\dot{f}_{\rm rot}$) from the ATNF pulsar catalog~\cite{ATNF_catalog}. To curate a clean and representative sample, we apply the selection cuts that canonical pulsars are removed via a rotational frequency cut ($f_{\rm rot} > 100~{\rm Hz}$)~\cite{Woan:2018tey}. 
Crucially, we filter the catalog to remove sources with a measured spin-up ($\dot{f}_{\rm rot} > 0$). Such sources are typically found in globular clusters, where local dynamical effects, such as acceleration within the cluster's gravitational potential, can induce an apparent spin-up that masks the intrinsic spin-down rate~\cite{Freire:2017mgu,KAGRA:2022dwb}. This filtering yields a cleaner sample from which to infer the true spin-down properties of the Galactic disk population. In version 2.6.5 of the ATNF pulsar catalog, a total of 673 MSPs satisfy the criterion $f_{\rm rot} > 100~{\rm Hz}$, of which 54 exhibit $\dot{f}_{\rm rot} > 0$. After excluding these sources, 619 MSPs remain and serve as the basis for Pop.~\Rmnum{1}.

To ensure greater precision, the Shklovskii effect~\cite{1970SvA....13..562S} and the differential Galactic rotation~\cite{Damour:1990wz} should be considered in correcting the $\dot{P}$ (also consequently $\dot{f}_{\rm rot}$). The GW detectability, discussed in the subsequent sections, mostly depends on the high-ellipticity tail of the ellipticity distribution, which scales with $\dot{P}$ (Eqs.~\eqref{eq_ellipticity_from_Bfield_scaling} and \eqref{eq_epsilon_from_eta}). However, as demonstrated in Fig.~1 of Ref.~\cite{Woan:2018tey}, such corrections are negligible for rather larger values of $\dot{P}$. Therefore, the influence of these effects on GW detectability is insignificant in our work.

\begin{figure*}[htbp]
\centering
\begin{subfigure}[t]{0.315\textwidth}
\includegraphics[width=\linewidth]{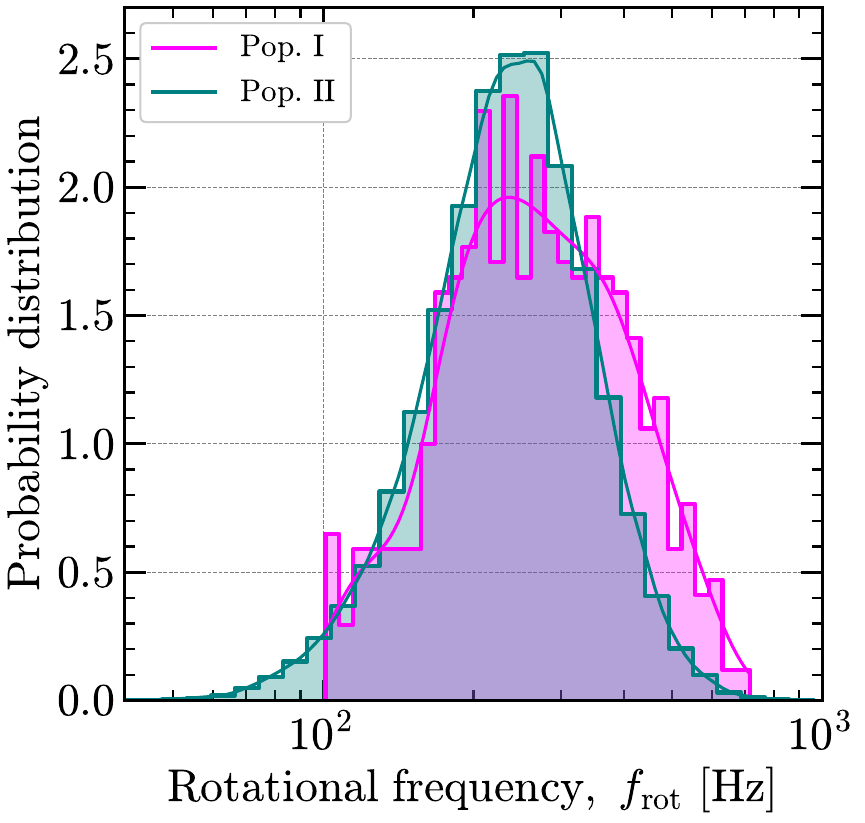}
\end{subfigure}
\begin{subfigure}[t]{0.33\textwidth}
\includegraphics[width=\linewidth]{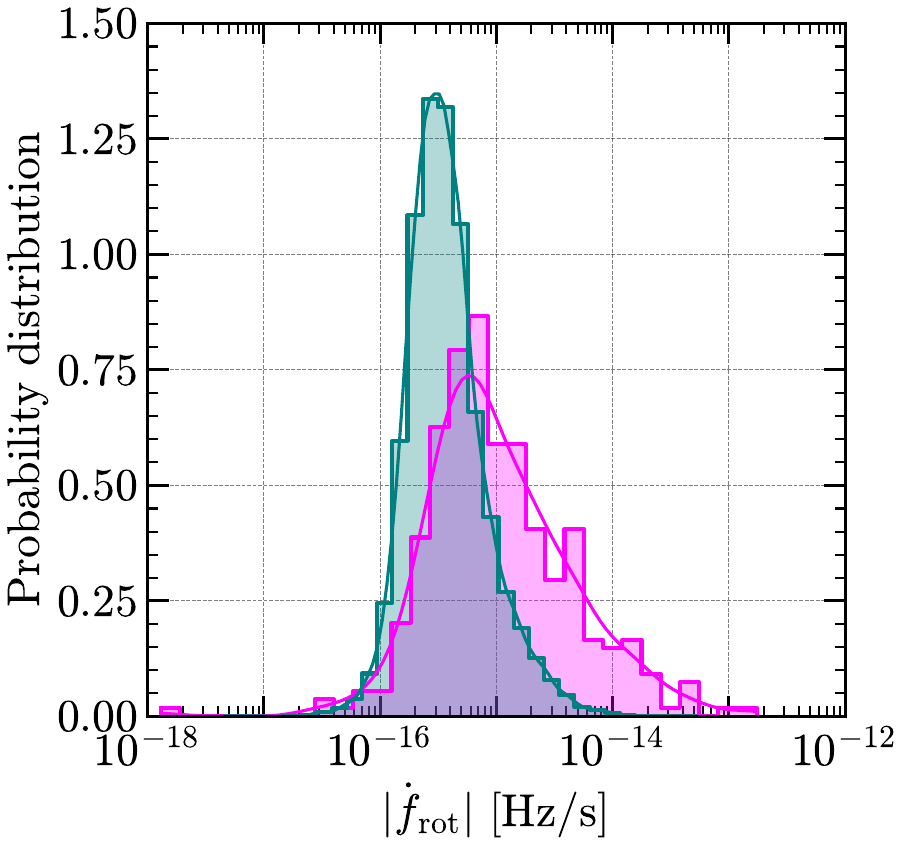}
\end{subfigure}
\begin{subfigure}[t]{0.305\textwidth}
\includegraphics[width=\linewidth]{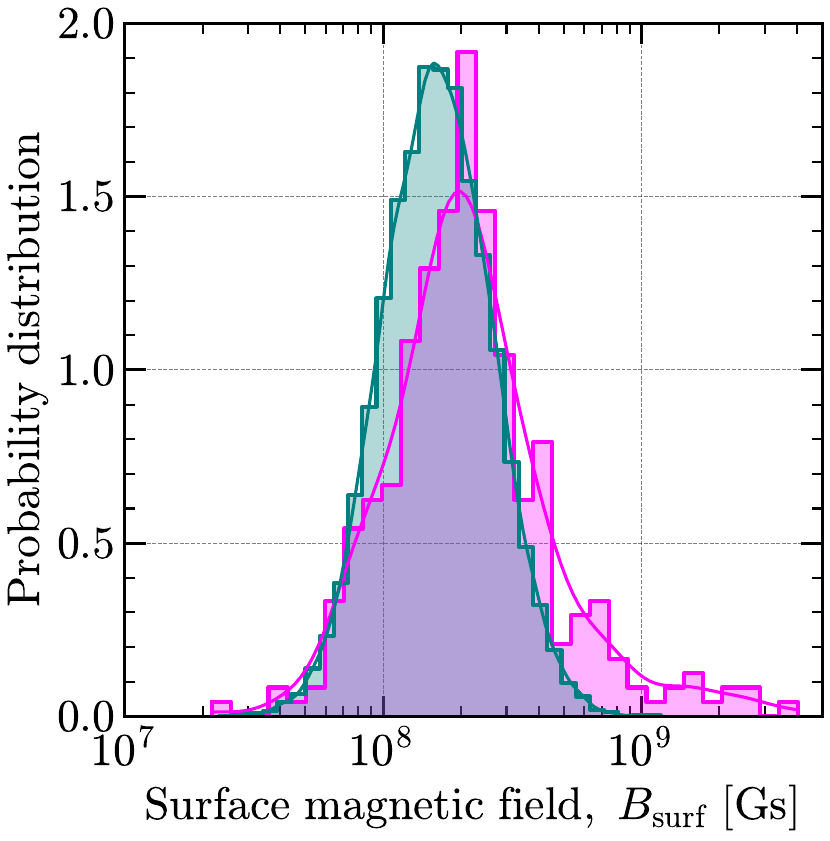}
\end{subfigure}
\caption[OptionalShort]{\justifying
Distributions of parameters of bulge MSPs responsible for the GCE.
\textcolor{magenta}{\textbf{Pink:} Pop.~\Rmnum{1}} derived from the filtered ATNF pulsar catalog~\cite{ATNF_catalog}.
\textcolor{teal}{\textbf{Green:} Pop.~\Rmnum{2}} derived from Ploeg et al.~\cite{Ploeg_2020}.
\textbf{Left and Middle}: Distributions of the rotational frequency $f_{\rm rot}$ and its derivative $\dot{f}_{\rm rot}$, respectively.
\textbf{Right}: Distributions of the surface magnetic field $B_{\rm surf}$, computed using Eq.~\eqref{eq_Bsurf}. In all panels, the curves represent the KDEs derived from the sample for the corresponding parameters. }
\label{fig_para_dist}
\end{figure*}

From this selected sample, we generate smooth probability density functions (PDFs) for the observable quantities $f_{\rm rot}$ and $\dot{f}_{\rm rot}$ using a kernel density estimation (KDE) method~\footnote{The KDE method can be found in the documentation: \url{https://docs.scipy.org/doc/scipy/reference/generated/scipy.stats.gaussian_kde.html}.}. 
The distribution for the surface magnetic field, $B_{\rm surf}$, is then derived following Eq.~\eqref{eq_Bsurf}.
These three resulting PDFs collectively define the present-day parameter distributions of Pop.~\Rmnum{1}, as presented in Fig.~\ref{fig_para_dist}.

For our second framework, Pop.~\Rmnum{2}, we generate present-day parameter distributions from a set of universal birth parameters, following the methodology and best-fit results of Ploeg et al.~\cite{Ploeg_2020}. 
Reference~\cite{Ploeg_2020} provides the necessary birth-property distributions: a log-normal distribution for the surface magnetic field, $B_{\rm surf}$, and another for the initial rotation period, $P_i$. We then evolve these initial periods to their present-day values using the age distribution of the bulge MSPs, which is derived from a convolution of the bulge's ancient star formation rate and the delay time distribution. We note that Ref.~\cite{Ploeg_2020} models the bulge as two components: a larger, dominant boxy bulge and a smaller nuclear bulge. Their best-fit results indicate that the number of MSPs in the boxy bulge is substantially larger than in the nuclear bulge (by a factor of about 5). Therefore, for our model, we simplify this by considering only the dominant boxy bulge component, using its specific star-formation history to derive the age distribution.

By sampling from the birth distributions of $P_i$ and $B_{\rm surf}$ and evolving each MSP over its sampled age, we generate the present-day period ($P$) distribution for the bulge population~\footnote{See Eq.~(2.23) and the surrounding text in Ref.~\cite{Ploeg_2020}.}. With the distributions for the present-day period $P$ and the (time-independent) magnetic field $B_{\rm surf}$ established, we then induce the distribution for the period derivative, $\dot{P}$, using the same force-free magnetosphere relation as before, as follows:
\be
    \dot{P} = \frac{5 \pi^2 R_{\rm MSP}^6 B_{\rm surf}^2}{3 c^3 I_{zz} P}.
\ee
By simulating a large ensemble, we produce the final evolved distributions for $f_{\rm rot}$, $\dot{f}_{\rm rot}$, converted from $P$ and $\dot{P}$, and $B_{\rm surf}$, presented in Fig.~\ref{fig_para_dist}. This provides a self-consistent, physically motivated model of the Galactic bulge MSP population.

\subsection{Ellipticity distribution}
\label{sec_pop_ellipticity}

\begin{figure*}[htbp]
\centering
    \begin{subfigure}[t]{0.32\textwidth}
        \includegraphics[width=\linewidth]{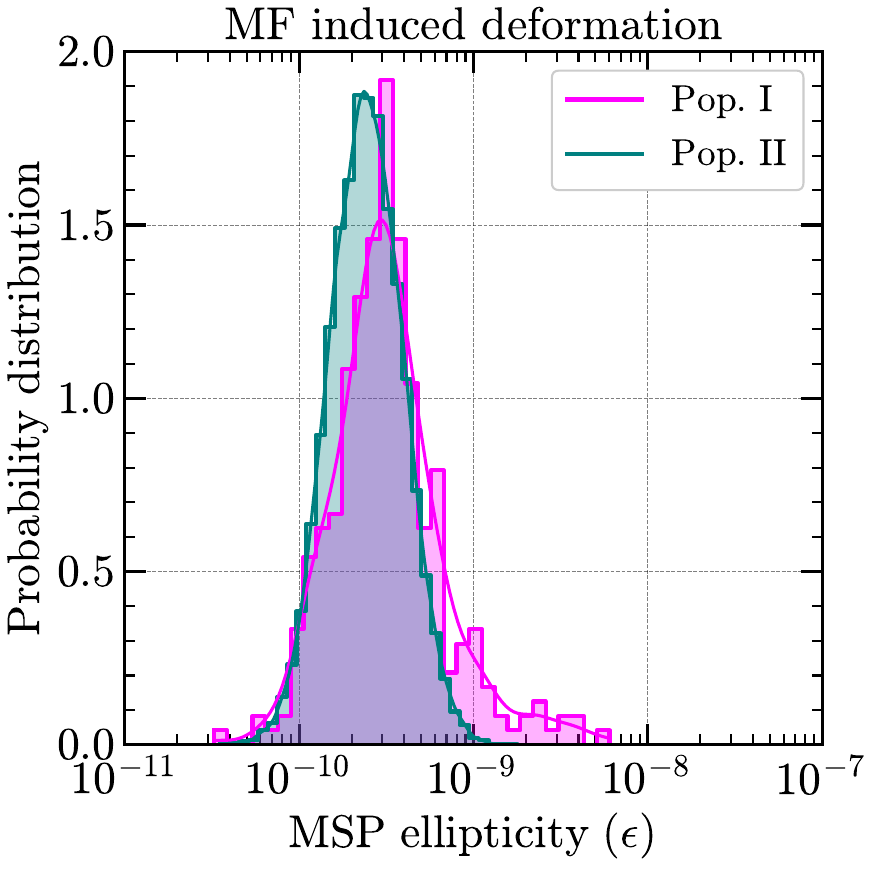}
    \end{subfigure}
    \begin{subfigure}[t]{0.32\textwidth}
        \includegraphics[width=\linewidth]{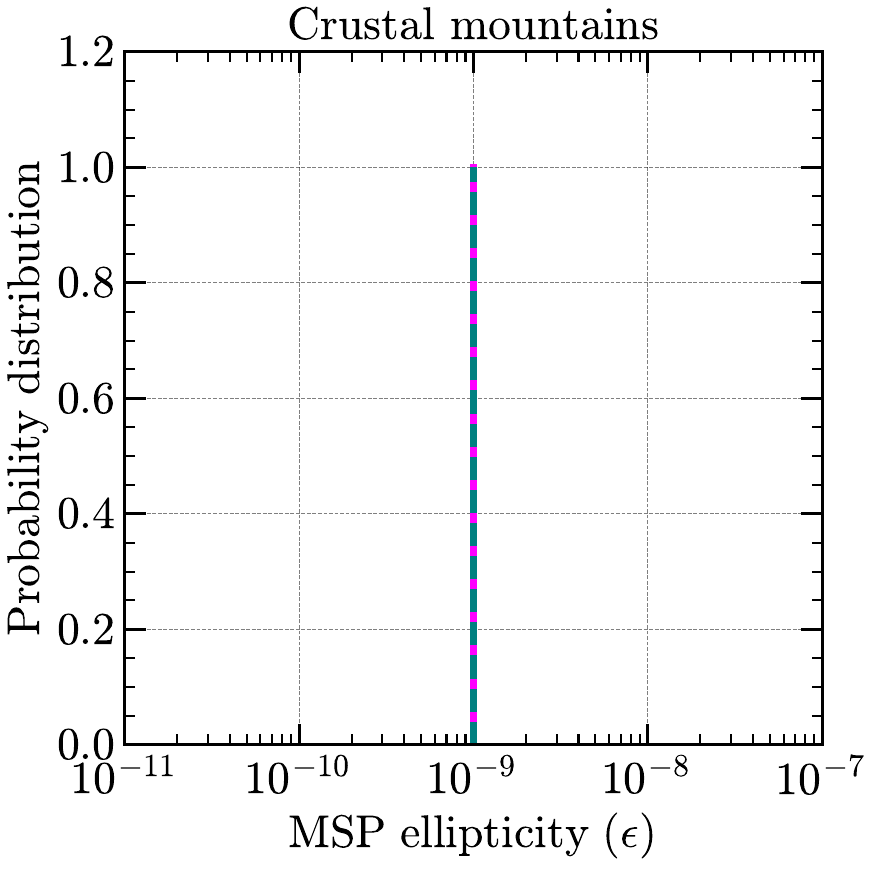}
    \end{subfigure}
    \begin{subfigure}[t]{0.33\textwidth}
        \includegraphics[width=\linewidth]{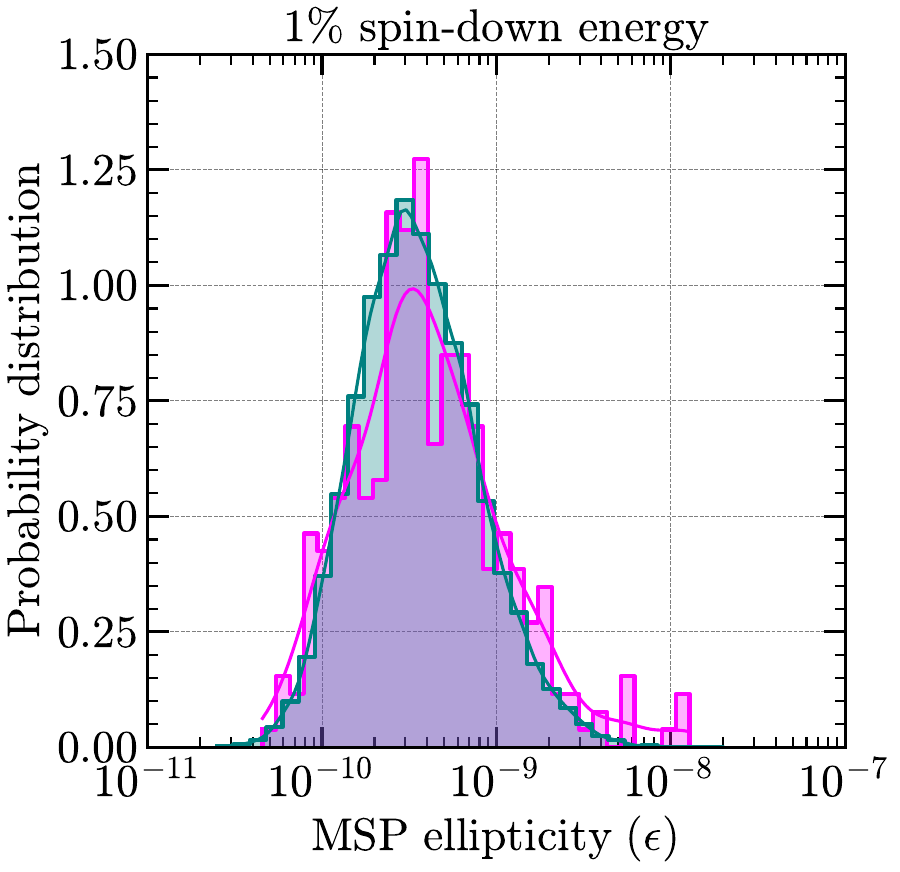}
    \end{subfigure}
\caption[OptionalShort]{\justifying
Distributions of ellipticities of bulge MSPs from three distinct scenarios.
\textcolor{magenta}{\textbf{Pink:} Pop.~\Rmnum{1}} derived from the filtered ATNF pulsar catalog~\cite{ATNF_catalog}.
\textcolor{teal}{\textbf{Green:} Pop.~\Rmnum{2}} derived from Ploeg et al.~\cite{Ploeg_2020}. 
\textbf{Left:} Magnetic-field-induced deformation. 
\textbf{Middle:} Crust mountains from breaking strain. 
\textbf{Right:} Model independent spin-down energy fraction with $\eta=1\%$. 
The curves in the left and right panels represent the KDEs derived from the sample for the corresponding parameters. } 
\label{fig_ellipticity_dist}
\end{figure*}

Fig.~\ref{fig_ellipticity_dist} presents the ellipticity distributions of bulge MSPs under the three considered scenarios. These distributions are obtained by combining the parameter distributions from the previous subsection with the ellipticity origins discussed in Sec.~\ref{sec_ellipticity}. For the magnetic-field-induced deformation scenario (left panel), Pop.~\Rmnum{1} yields slightly higher ellipticities than Pop.~\Rmnum{2}, reflecting their differing distributions of $B_{\rm surf}$, $f_{\rm rot}$, and $|\dot{f}_{\rm rot}|$ (Fig.~\ref{fig_para_dist}). 
For the crustal mountains scenario (middle panel), both populations exhibit the same $\delta$-functionlike distribution. 
For the energy-loss-fraction scenario (right panel), the two populations again show nearly identical distributions, as differences in $f_{\rm rot}$ and $|\dot{f}_{\rm rot}|$ effectively cancel out in Eq.~\eqref{eq_epsilon_scaling_Eloss}.

\subsection{Morphology}
\label{sec_pop_morphology}

The morphology of the GCE provides crucial information about its origin. While early analyses suggested a spherically symmetric profile, such as a generalized Navarro-Frenk-White (gNFW) profile consistent with DM annihilation, compelling recent evidence indicates that the GCE's spatial distribution may correlate with the stellar mass of the Galactic bulge~\cite{Bartels:2015aea, Lee:2015fea, Macias:2016nev, Bartels:2017vsx}. 
This finding thus supports an astrophysical origin, such as a population of unresolved MSPs~\footnote{See also Ref.~\cite{Muru:2025vpz} for a recent study, which shows that realistic MW simulations yield a flattened, boxy DM distribution, making the GCE morphology from WIMP annihilation hard to distinguish from that of MSPs.}.

Given our focus on investigating the GW signal from the MSPs responsible for the GCE, we adopt a morphology that traces the stellar bulge over a gNFW profile. Following the physically motivated model in Refs.~\cite{Macias:2016nev,Bartels:2017vsx,Ploeg_2020}, we specifically adopt the boxy bulge configuration~\cite{1998ApJ...492..495F,Macias_2019}. 
We consider another well-motivated bulge model---the spherical bulge in Appendix~\ref{sec_app_morpho_dep}, and find that \textit{our GW signal predictions are insensitive to the assumed bulge morphology}.

This spatial distribution of the boxy bulge model is described by
\begin{equation}
    \rho_{\rm boxy} \propto {\rm sech}^2(R_s) \times
    \begin{cases}
        1 & R \leq R_{\rm end} \\ \exp \left\{ -\frac{(R - R_{\rm end})^2}{h_{\rm end}^2} \right\} & R > R_{\rm end},
    \end{cases}
\label{boxy_bulge_rho}
\end{equation}
where $R$ is the cylindrical Galactocentric radial distance, with $R_{\rm end}=3128$~pc and $h_{\rm end} = 461$~pc. The quantity $R_s$ is a scaled radius defined by
\begin{align}
    R_{\perp}^{C_\perp} &= \left( \frac{|x'|}{1696 ~ {\rm pc}} \right)^{C_\perp} + \left( \frac{|y'|}{642.6 ~ {\rm pc}} \right)^{C_\perp} \\
    R_s^{C_\parallel} &= R_{\perp}^{C_\parallel} + \left( \frac{|z'|}{442.5 ~ {\rm pc}} \right)^{C_\parallel},
\label{boxy_R_prep_S}
\end{align}
with power-law indices $C_\parallel = 3.501$ and $C_\perp = 1.574$. The primed coordinates ($x', y', z'$) are Cartesian coordinates within the boxy bulge's reference frame. This frame is obtained by rotating the standard Galactocentric frame (in which the Sun is located at $x_\odot = -R_0, y_\odot=z_\odot=0$) first by $13.79^\circ$ around the $z$-axis, and then by $0.023^\circ$ around the new $y'$-axis.

\subsection{MSPs' total number and luminosity functions}
\label{sec_pop_NMSP}

For Pop.~\Rmnum{1}, the total population size, $N_{\rm MSP}$, is a key input parameter that must be benchmarked from existing literature. The comprehensive analysis of Ref.~\cite{Dinsmore:2021nip} provides an ideal framework for this purpose. This work tested several benchmark luminosity functions, finding that models based on globular cluster or non-Poissonian template fitting data are in strong tension with observations, as they overproduce the number of resolved point sources. From the observationally viable frameworks presented in their analysis, we select the accretion-induced collapse (AIC)~\cite{Gautam:2021wqn} benchmark as the basis for our population estimate. This choice is dually motivated: first, the model is fully consistent with the resolved source constraints from the GCE. Second, its prediction, a total population of $N_{\rm gNFW} \approx 3.6 \times 10^5$ for a gNFW profile~\cite{Navarro:1995iw,Zhao:1995cp,Navarro:1996gj}, aligns with independent, recent analyses using machine learning, which show a preference for a large population of $N_{\rm MSP} > 10^5$~\cite{List:2025qbx}.

Since our model adopts the boxy bulge morphology, we must rescale this benchmark number to account for the different spatial geometry while preserving the total GCE flux. The scaling is determined by the ratio of the flux-to-luminosity factors ($F/L$) for the two profiles. We numerically compute the $F/L$ ratio for both the gNFW and our boxy bulge profiles using the standard relation
\begin{equation}
\frac{F}{L} = \frac{1}{4\pi} \frac{\int_{\Omega}d\Omega\int_0^{\infty}ds\,\rho(s,\Omega)}{\int_{\Omega}d\Omega\int_0^{\infty} s^2 ds\,\rho(s,\Omega)},
\label{equation of F/L}
\end{equation}
where $\rho$ is the spatial density for the respective profile. The required number of MSPs for our model is then given by the scaling relation:
\begin{equation}
    N_{\rm boxy} = N_{\rm gNFW} \times \frac{(F/L)_{\rm gNFW}}{(F/L)_{\rm boxy}}.
\end{equation}
Our calculation of this geometric factor indicates that for our region of interest (ROI), the boxy bulge profile is slightly more spatially concentrated than the gNFW profile, resulting in a scaling factor of approximately 0.8. Applying this factor to the AIC benchmark number, we obtain our baseline estimate of approximately $2.9 \times 10^5$ MSPs for Pop.~\Rmnum{1}.

In contrast to Pop.~\Rmnum{1}, the total size of the bulge population for Pop.~\Rmnum{2} is not a free parameter but a derived quantity. It is determined by the normalization required for the model's derived luminosity function to reproduce the total GCE flux. We therefore adopt the total population size as estimated by Ploeg et al.~\cite{Ploeg_2020}, but must rescale it to our specific ROI.

Their analysis, which fits the GCE flux within the region $|\ell| < 20^{\circ}$ and $|b| < 20^{\circ}$ (hereafter $\Omega_{\rm Ploeg}$), suggests a total of $N_{\rm Ploeg} \approx 34000$ MSPs. This number, however, represents the combined population from both the boxy and nuclear bulge components described in their model. Our analysis focuses on the boxy bulge, as our ROI mask ($|b|>2^\circ$) preferentially removes the central Galactic plane where the highly concentrated nuclear bulge resides. We therefore first isolate the number of MSPs belonging to the boxy bulge component. Using the best-fit ratio from Ref.~\cite{Ploeg_2020}, $\log_{10}(N_{\rm nuclear}/N_{\rm boxy}) \approx -0.66$, we can get the boxy bulge comprises a fraction $f_{\rm boxy} \approx 0.82$ of the total population.

It is also necessary to account for the differences in the solid angles of the analysis regions. We calculate the fraction of the boxy bulge population residing within our masked ROI ($|\ell| < 20^{\circ}$, $2^{\circ} < |b| < 20^{\circ}$, hereafter $\Omega_{\rm ROI}$) relative to the population within the unmasked region of Ref.~\cite{Ploeg_2020}. This geometric scaling factor, $f_{\rm ROI}$, is the ratio of the volume integrals of the density profile over the respective solid angles, as follows:
\begin{equation}
    f_{\rm ROI} = \frac{\int_{\Omega_{\rm ROI}}d\Omega\int_0^{\infty}ds\,s^2\,\rho_{\rm boxy}(s,\Omega)}{\int_{\Omega_{\rm Ploeg}}d\Omega\int_0^{\infty}ds\,s^2\,\rho_{\rm boxy}(s,\Omega)}.
    \label{eq:roi_fraction}
\end{equation}
By numerically evaluating this expression, we find that approximately 58\% of the boxy bulge MSPs within $\Omega_{\rm Ploeg}$ are contained in our ROI. The final population size for our model is therefore the product of these two correction factors applied to the total population range from Ref.~\cite{Ploeg_2020}, $N_{\rm MSP} = N_{\rm Ploeg} \times f_{\rm boxy} \times f_{\rm ROI}$. This yields a baseline estimate of $\mathbf{N_{\rm MSP} \approx 16170}$ for Pop.~\Rmnum{2}.

\section{MSP GW detection strategies: Coherent versus Incoherent}
\label{sec_detect}

\subsection{Continuous Monochromatic GW Detection}
\label{sec_detect_simplest}

As shown in Eq.~\eqref{eq_h}, the GW emission from an MSP is monochromatic, at $f_{\rm GW} = 2 f_{\rm rot}$. 
In this case, the minimum detectable strain amplitude for a continuous monochromatic wave at a given signal-to-noise ratio (SNR) is~\cite{Maggiore:2007ulw}
\begin{equation}
h_0^{\min} = 
\frac{{\rm SNR}}{\left\langle F_{+}^2\right\rangle^{1 / 2}} \sqrt{\frac{S_n\left(f_{\rm GW}\right)}{T_{\rm obs}}} \cdot \frac{1}{\sqrt{4/5}}\,,
\label{eq_h0min_coh}
\end{equation}
where $S_n(f_{\rm GW})$ represents the power spectrum density (PSD) of detectors evaluated at frequency $f_{\rm GW}$, and $T_{\rm obs}$ is the effective observation time. 
For ${\left\langle F_{+}^2\right\rangle^{1/2}}$, 
we use the root-mean-square antenna response to the ``$+$'' polarization, averaged over source sky position and polarization during one sidereal day, yielding $\sqrt{1/5}$ for interferometers. 
The extra factor $\sqrt{4/5}$ arises from averaging over an isotropic distribution of the inclination angle $\iota$ between the pulsar rotation axis and the line of sight, as described in Eq.~\eqref{eq_h}.
Finally, for our MSPs clustered near the GC rather than isotropically distributed over the full sky, the antenna pattern average differs from the all-sky value by only a few percent~\cite{KAGRA:2022osp, Miller:2023qph}, which we neglect.

The frequency resolution is given by
\be
(\Delta f)^{\rm res} \simeq \frac{1}{T_{\rm obs}} \,.
\ee
The decrease of $h_0^{\min}$ with $T_{\rm obs}$ in Eq.~\eqref{eq_h0min_coh} can also be understood as that reducing the bandwidth ($1/T_{\rm obs}$) accepts less noise while the monochromatic signal stays the same.

\subsection{Complications due to Doppler Shifts}
\label{sec_detect_doppler}

However, the signal at detection at a specific time is shifted in frequency due to Doppler effects, which originate from the Earth's rotation and revolution and possible orbital motion of the MSP if it is in a binary system. In a time period $T_{\rm obs}$, the relative velocity of the Earth and the source changes with time, which produces a time-varying Doppler shift in the frequency. 
In the frequency domain, this manifests as a Doppler broadening/smearing, with the width set by the maximum Doppler shift with $T_{\rm obs}$, denoted by $(\Delta f)^{\rm max}$. Once the $(\Delta f)^{\rm max}$ becomes comparable to the frequency resolution $1/T_{\rm obs}$, increasing $T_{\rm obs}$ further does not improve substantially the SNR, as it reduces both accepted noise and signal, and Eq.~\eqref{eq_h0min_coh} is no longer applicable.

We estimate the maximum Doppler shifts, $(\Delta f)^{\rm max}$, as follows:
first, considering the Earth's rotation, at latitude $40^\circ$ (approximately matching the location of the current and future ground-based GW detectors), 
the surface velocity is 
$v_{\text {rot }}=\omega_{\text {rot }} R_{\oplus} \cos \left(40^{\circ}\right) \simeq 355 \mathrm{~m} / \mathrm{s}$, where $\omega_{\rm rot} = 2\pi/(24\, {\rm hr})$ and $R_{\oplus}$ is the Earth's radius. 
This leads to a maximum frequency shift when the detector reverses its velocity relative to the MSP (every $\simeq12$ hours)~\cite{Maggiore:2007ulw}, as follows:
\be
(\Delta f)^{\max}_{\mathrm{rot}} 
\simeq 
2 f_0 \frac{v_{\mathrm{rot}}}{c} 
\simeq 
4.8\times 10^{-3} \mathrm{~Hz}  \left(\frac{f_0}{ 2 \, \mathrm{kHz}} \right) \,.
\ee
Second, for Earth's revolution (approximated as circular with $v_{\rm{orb}} \simeq 3 \times 10^4 \, \rm{m/s}$), the maximum Doppler shift is (every $\simeq6$ months)~\cite{Maggiore:2007ulw}
\begin{equation}
(\Delta f)^{\max }_{\mathrm{orb}} \simeq 2 f_0 \frac{v_{\text {orb }}}{c} \simeq 0.4 \mathrm{~Hz}\left(\frac{f_0}{2 \, \mathrm{kHz}}\right).
\end{equation}
Finally, for the orbital motion of the MSPs in the binary systems, the maximum Doppler shift (every half orbital period),
as estimated from the ATNF pulsar catalog samples~\cite{ATNF_catalog}, is found to be
\begin{equation}
(\Delta f)^{\max }_{\mathrm{bin}} \simeq 2 f_0 \frac{v_{\text {bin }}}{c} \simeq 2.0 \mathrm{~Hz}\left(\frac{f_0}{2 \, \mathrm{kHz}}\right),
\label{eq_Deltaf_max_binary}
\end{equation}
with further details provided in Appendix~\ref{sec_app_doppler}.

Therefore, as long as $T_{\rm obs}$ is longer than $\sim 1$ hour, the Doppler shifts are important and Eq.~\eqref{eq_h0min_coh} is no longer applicable.

Doppler shift can, in principle, be corrected by defining a new time variable $t^{\prime}=t + ({\mathbf{x}(t) \cdot \hat{\mathbf{r}}})/{c}$~\footnote{See Eq.~(7.149) and the surrounding text in Ref.~\cite{Maggiore:2007ulw}.}, where $\mathbf{x}(t)$ is the position of the detector and $\hat{\mathbf{r}}$ is the unit vector pointing toward the source. However, to correct the Doppler shift, the location of the MSPs needs to be known accurately~\footnote{See Eq.~(7.151) and the surrounding text in Ref.~\cite{Maggiore:2007ulw}.}, e.g., 
\be
\Delta \theta < 0.1 \operatorname{arcsec}\left(\frac{10^7\, \mathrm{s}}{T}\right)^2\left(\frac{1\, \mathrm{kHz}}{f_0}\right) \,.
\label{eq_Doppler_DeltaTheta}
\ee
For bulge MSPs with unknown locations, one could in principle perform blind searches over the GCE region using sufficiently small pixel sizes. However, the sky-position requirement in Eq.~\eqref{eq_Doppler_DeltaTheta} is extremely demanding for blind Galactic Center searches, with the computational cost scaling at least as $\sim T_{\rm obs}^5$.
Furthermore, even if the Doppler shifts due to the Earth’s motion can be corrected by redefining the time variable, the orbital motion of bulge MSPs in binary systems remains unknown. Since most MSPs are expected to be in binaries, such unmodeled Doppler modulation cannot be corrected in the same way.

\subsection{Our Approach: Two Detection strategies}
\label{sec_detect_ourwork}

Motivated by the discussion above, we consider two strategies for detecting bulge MSPs. 

The first is the coherent strategy, which assumes that all Doppler shifts can be corrected. 
This is feasible for isolated MSPs in blind searches with sufficiently small pixel sizes (Eq.~\eqref{eq_Doppler_DeltaTheta}), provided that adequate computational power is available.
In this case, an individual MSP becomes detectable once its GW strain amplitude ($h_0$; Eq.~\eqref{eq_h0}) exceeds the sensitivity threshold ($h_0^{\min}$; Eq.~\eqref{eq_h0min_coh}) with ${\rm SNR} = 5$~\cite{KAGRA:2021una, LIGOScientific:2025kei} and $S_n$ for different detectors, i.e.,
\be
h_0 > h_0^{\min}\, .
\label{eq_cmpr_coh}
\ee

The second is the incoherent strategy, which assumes
that the MSP's Doppler shifts cannot be fully corrected,
either due to limited computational power in blind
searches or because of unknown orbital modulation in
binary systems. In this case, the signal from an individual
MSP is no longer confined to a single Fourier bin
($(\Delta f)^{\rm res} \sim 1/T_{\rm obs}$), but broadened to a width
$(\Delta f)^{\max} \gg 1/T_{\rm obs}$ set by the maximum Doppler
shift within $T_{\rm obs}$. 
As a consequence, a fully coherent integration over
the entire $T_{\rm obs}$ is not possible.

A standard approach is to divide the data into $N$
shorter coherent segments of duration $T_{\rm stack}$, such
that $1/T_{\rm stack} \sim (\Delta f)^{\max}$ holds within each segment.
Each stack is analyzed coherently, and then the results are
combined incoherently across $N = T_{\rm obs}/T_{\rm stack}$
segments. The corresponding sensitivity for incoherent
search is degraded compared to the coherent case by a
factor of $N^{1/4}$~\cite{Maggiore:2007ulw}, as follows:
\begin{equation}
h_0^{\min,{\rm incoh}} \;=\;
\frac{{\rm SNR}}{\langle F_+^2 \rangle^{1/2}}
\sqrt{\frac{S_n(f_0)}{T_{\rm obs}}} \cdot
\frac{1}{\sqrt{4/5}} \cdot 
\,\left( \frac{T_{\rm obs}}{T_{\rm stack}} \right)^{\frac{1}{4}},
\label{eq_h0min_incoh}
\end{equation}
where $(T_{\rm obs}/T_{\rm stack})^{1/4}$ quantifies the
loss of sensitivity due to loss of phase coherence across
the broadened bandwidth. Motivated by discussions in Sec.~\ref{sec_detect_doppler}, we choose $1/T_{\rm stack} = (\Delta f)^{\max} = 2$~Hz.
In the frequency domain, this is equivalent to binning the signals of the MSPs into different frequency windows. 

For $\langle F_+^2 \rangle$ in Eq.~\eqref{eq_h0min_incoh}, we adopt the standard sky-, polarization-, and inclination-averaged value of $1/5$, consistent with Eq.~\eqref{eq_h0min_coh}. Strictly speaking, such an average is valid when the observation time is long compared to a sidereal day, so that the apparent motion of the sources averages over right ascension and polarization.
In our case, although each coherent segment is only $T_{\rm stack}=0.5$~s, the total observation time can exceed a year. Moreover, the signal is an incoherent sum over $\mathcal{O}(10^5)$ MSPs near the GC (discussion below), whose orientations $(\iota,\psi)$ can be regarded as effectively random. These two facts justify the use of the averaged antenna pattern in Eq.~\eqref{eq_h0min_incoh}, and deviations from the exact instantaneous values are expected to be negligible in our setup.

Since the phase information is lost between different stacks, the incoherent strategy is sensitive only to the summed
power of the signals within the frequency window
$ (\Delta f)^{\max}$. In practice, the effective strain amplitude from multiple MSPs that fall into the same frequency window can be
written as
\begin{equation}
h_{\rm eff} \;=\;
\left(\frac{\pi}{4}\sum_{i \in (\Delta f)^{\max}} h_{0,i}^2 \right)^{1/2},
\label{eq_heff}
\end{equation}
where $h_{0,i}$ is the intrinsic strain amplitude of the $i$th MSP and the factor $\pi/4$ originates from the random selection of phase angles.
Detection occurs when
\begin{equation}
h_{\rm eff} \; > \; h_0^{\min,{\rm incoh}}.
\label{eq_cmpr_incoh}
\end{equation}

Therefore, in the incoherent strategy, the observables
are no longer individual monochromatic signals, but rather, the collective excess power within each Doppler
broadened frequency window~\footnote{This is somewhat analogous to the stochastic GW background~\cite{Allen:1997ad, Sharma:2020btq, Biscoveanu:2020gds, KAGRA:2021kbb, Mukherjee:2021itf, Zhou:2022otw, Zhou:2022nmt, Zhong:2022ylh, Zhong:2024dss, Zhong:2025qno}, though the MSP signals mainly from the galactic center.}. 
The GW detectability in this case scales as $h_{\rm eff} / h_0^{\min,{\rm incoh}} \propto ((\Delta f)^{\max})^{1/4}$. We adopt $(\Delta f)^{\max}=2$~Hz as a conservative choice.

Between the (fully) coherent and (fully) incoherent strategies discussed above, many continuous GW search pipelines employ a semicoherent approach~\cite{LIGOScientific:2013wcb,Piccinni:2019zub,Covas:2022rfg, KAGRA:2022osp, KAGRA:2022dwb,LIGOScientific:2026plm}. 
This allows exploring a large parameter space at a reasonable computational cost, with only a relatively small sensitivity loss compared to the coherent strategy.
In this case, the data are divided into segments that are moderately longer than $T_{\rm stack} = 0.5$ s used in our incoherent strategy, so that more phase information is preserved. Each segment is analyzed coherently, and the results are then combined incoherently. 
As a result, the sensitivity and detectability of semicoherent searches fall between the coherent and incoherent strategies considered above.

\section{Results and discussion}
\label{sec_results}

\newcommand{\subfigwidth}{0.39\textwidth}
\newcommand{\rightpanelwidth}{0.95\linewidth}

\begin{figure*}
\centering
\begin{subfigure}[t]{\subfigwidth}
    \includegraphics[width=\linewidth]{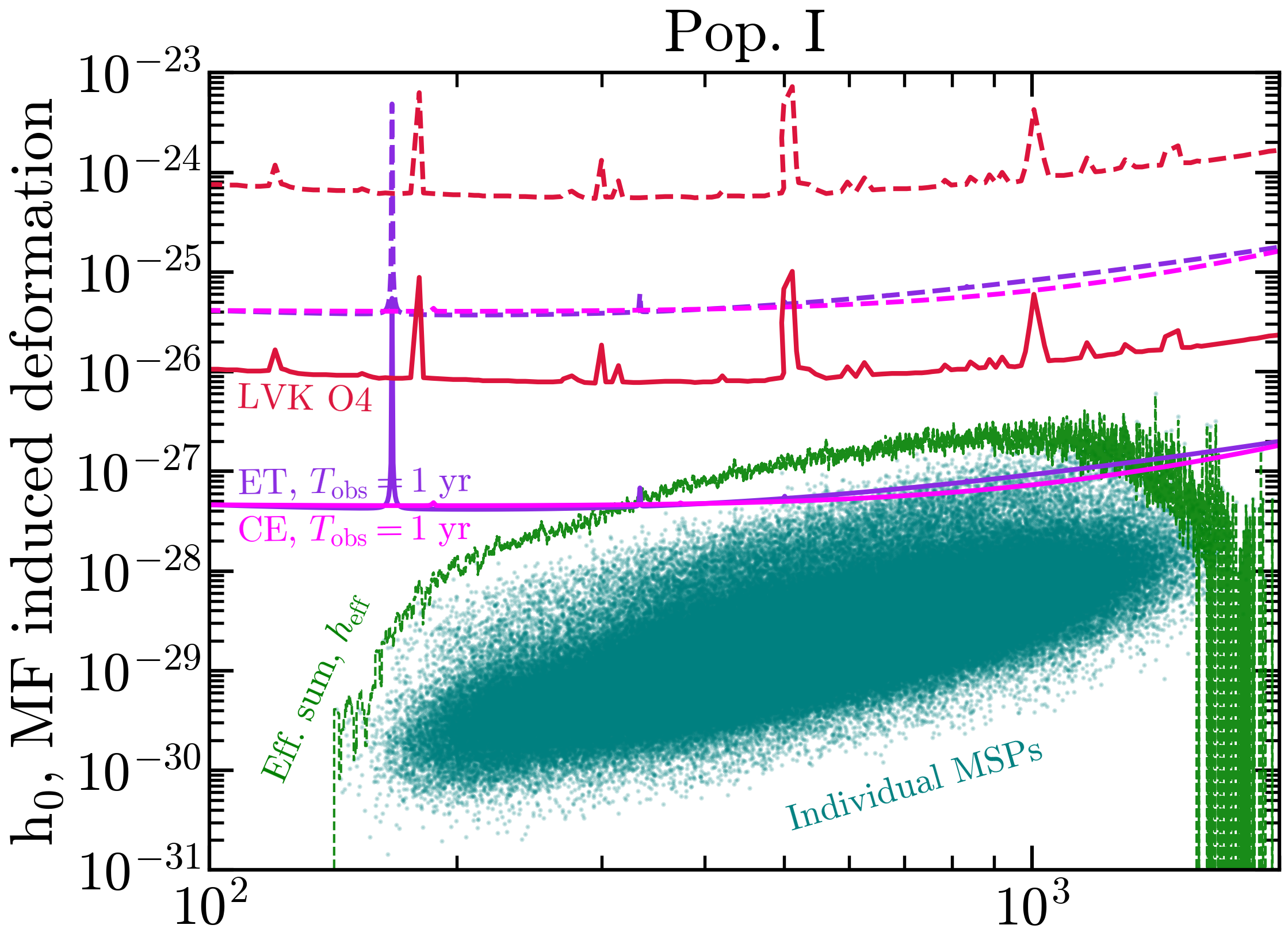}
\end{subfigure}
\begin{subfigure}[t]{\subfigwidth}
    \includegraphics[width=\rightpanelwidth]{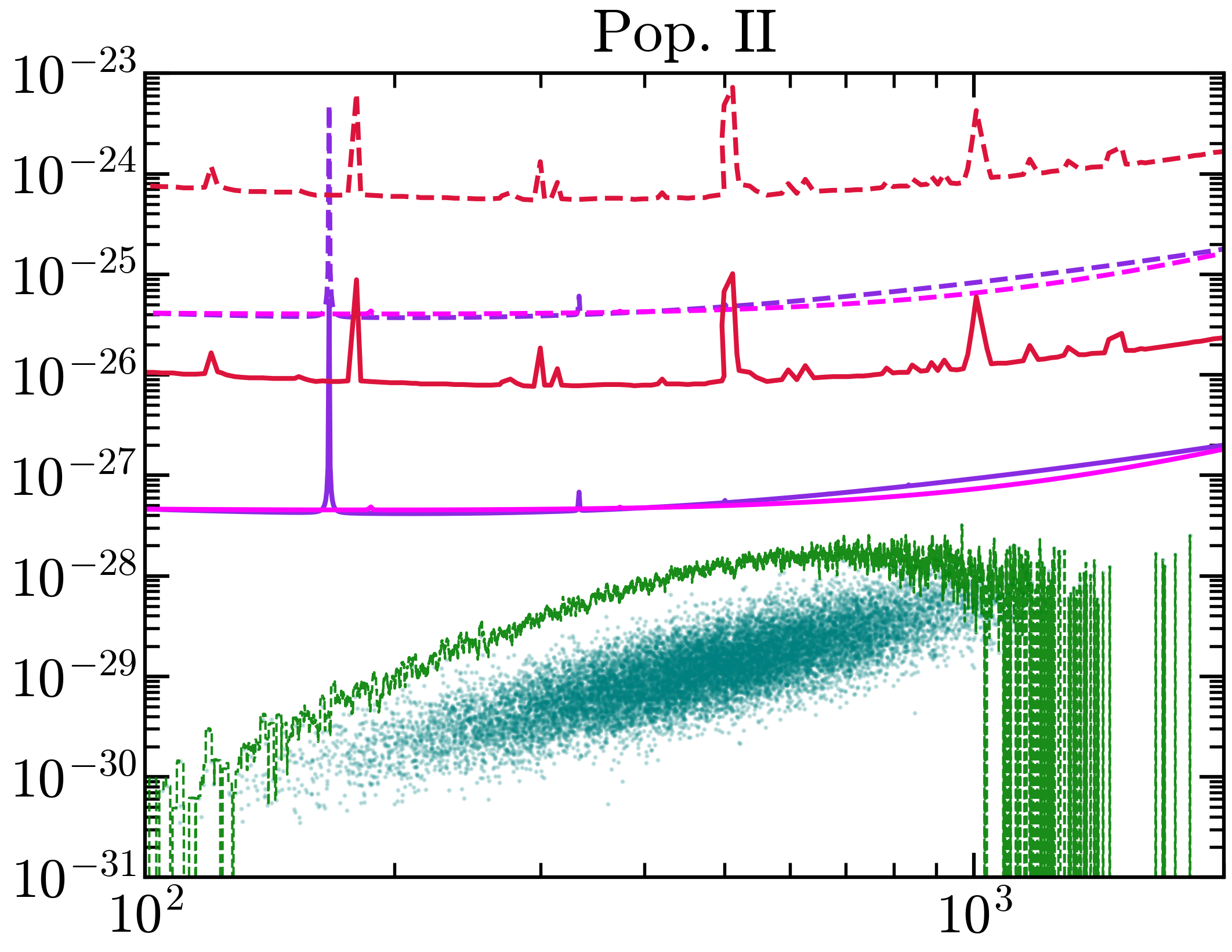}
\end{subfigure}
\\
\centering
\begin{subfigure}[t]{\subfigwidth}
    \includegraphics[width=\linewidth]{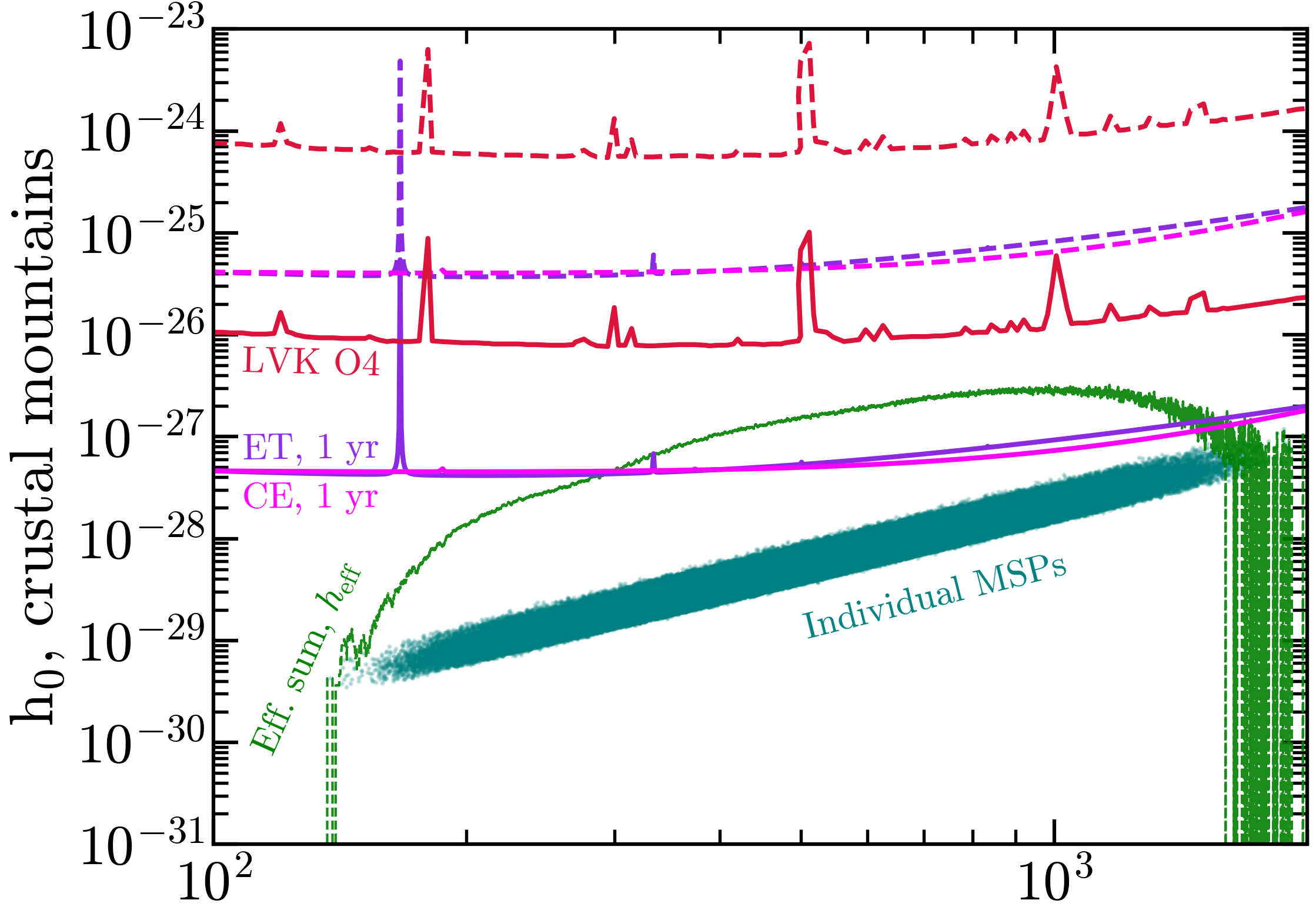}
\end{subfigure}
\begin{subfigure}[t]{\subfigwidth}
    \includegraphics[width=\rightpanelwidth]{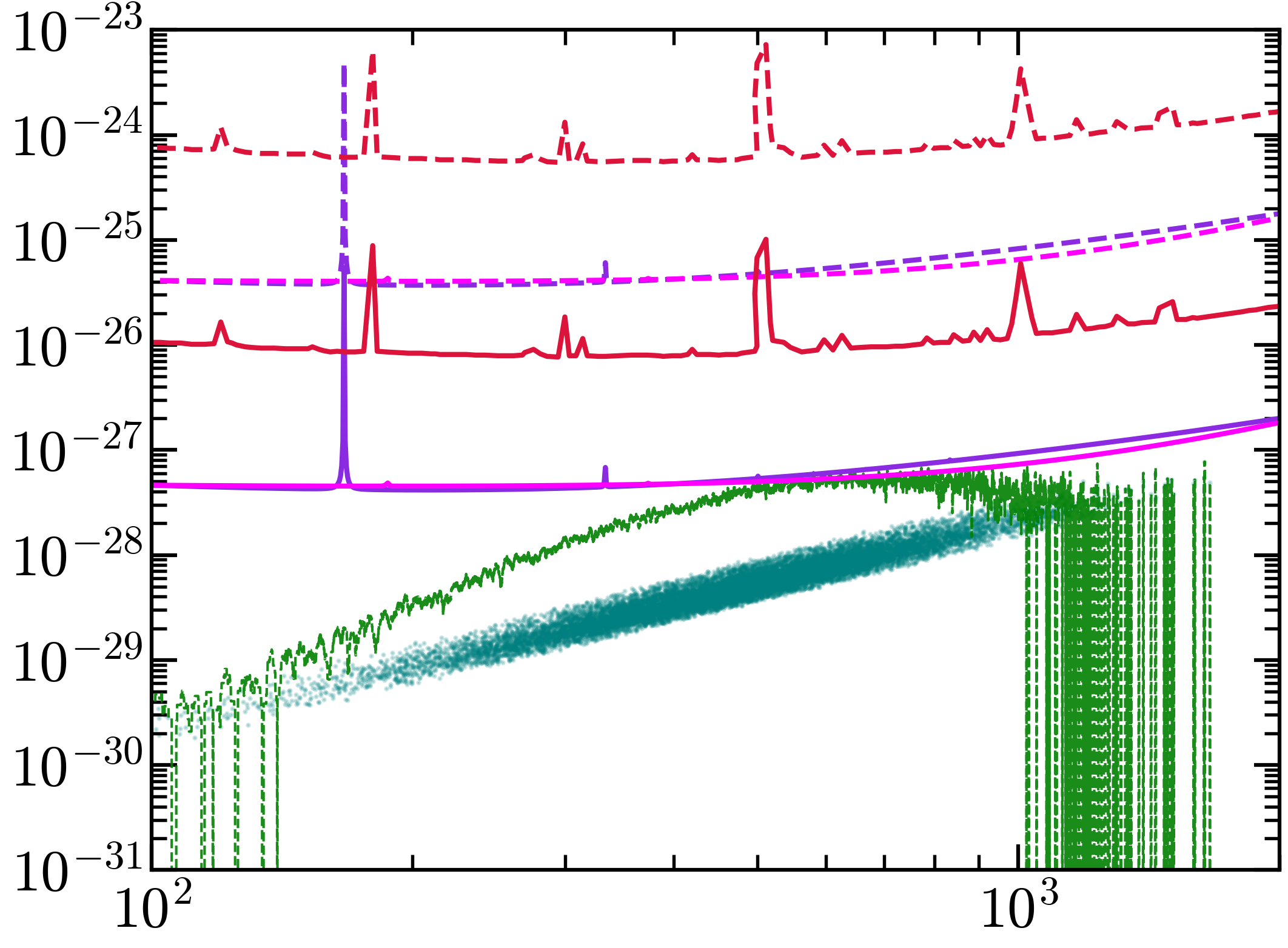}
\end{subfigure}
\\
\centering
\begin{subfigure}[t]{\subfigwidth}
    \includegraphics[width=\linewidth]{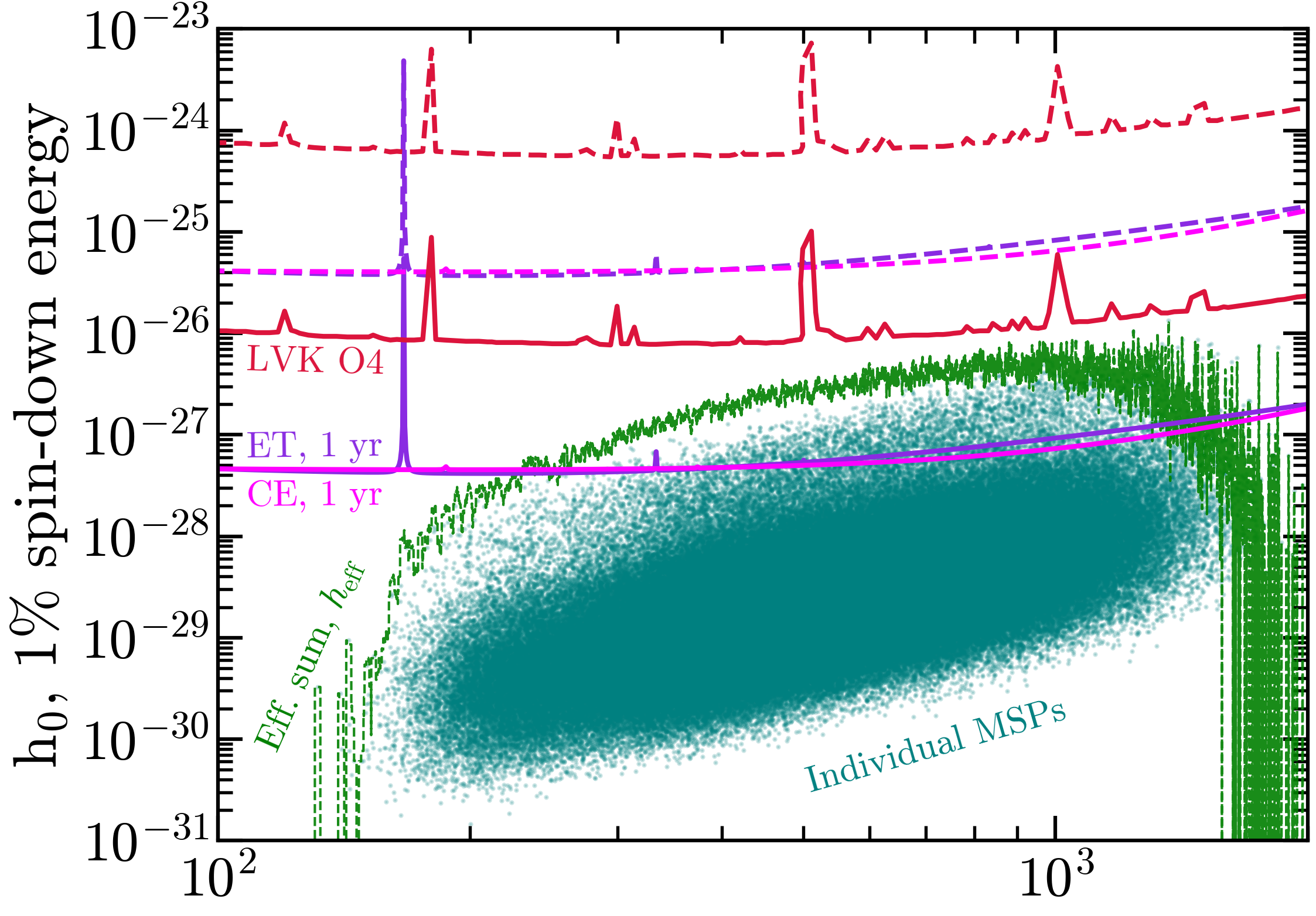}
\end{subfigure}
\begin{subfigure}[t]{\subfigwidth}
    \includegraphics[width=\rightpanelwidth]{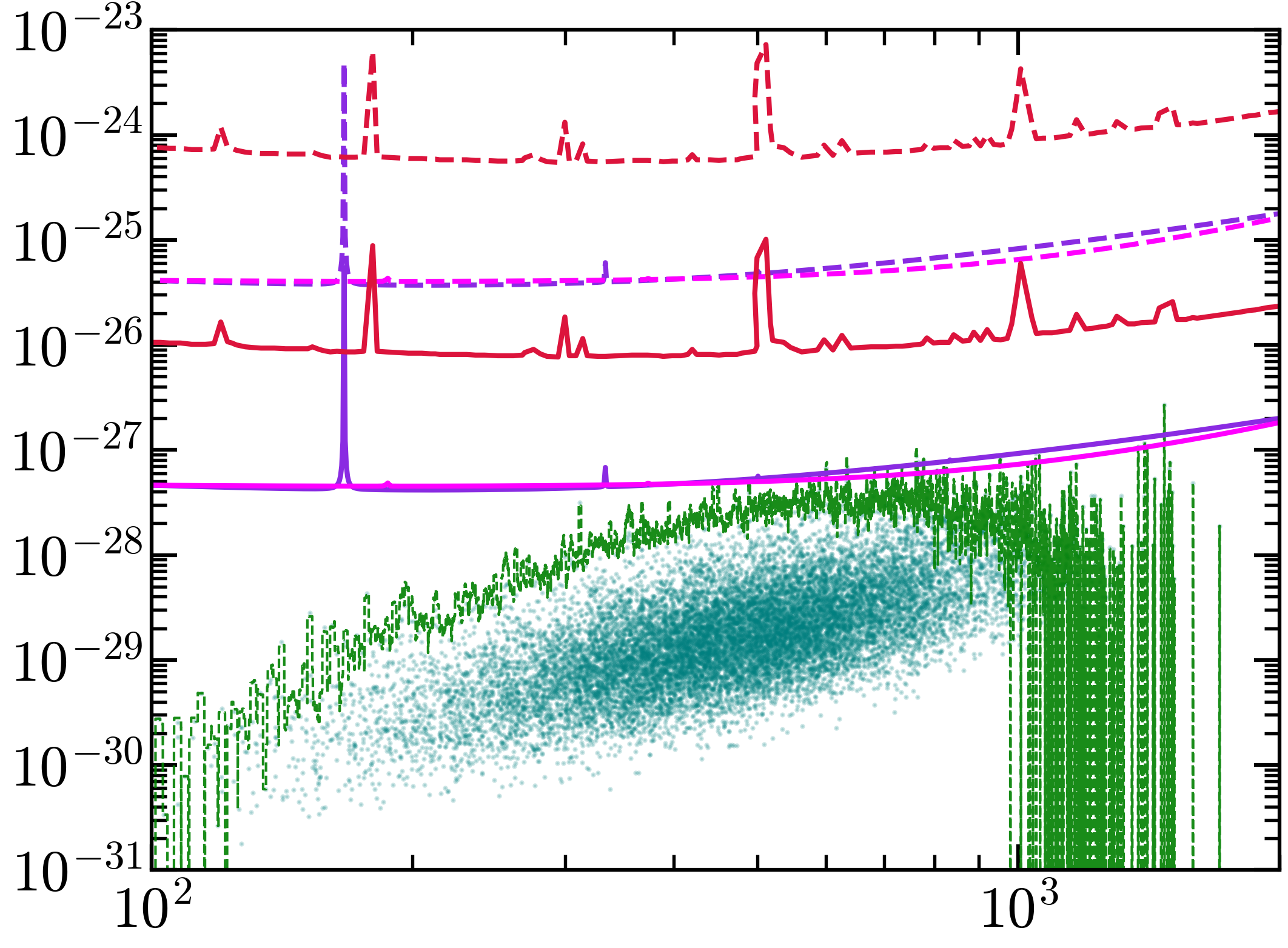}
\end{subfigure}
\\
\centering
\begin{subfigure}[t]{\subfigwidth}
    \includegraphics[width=\linewidth]{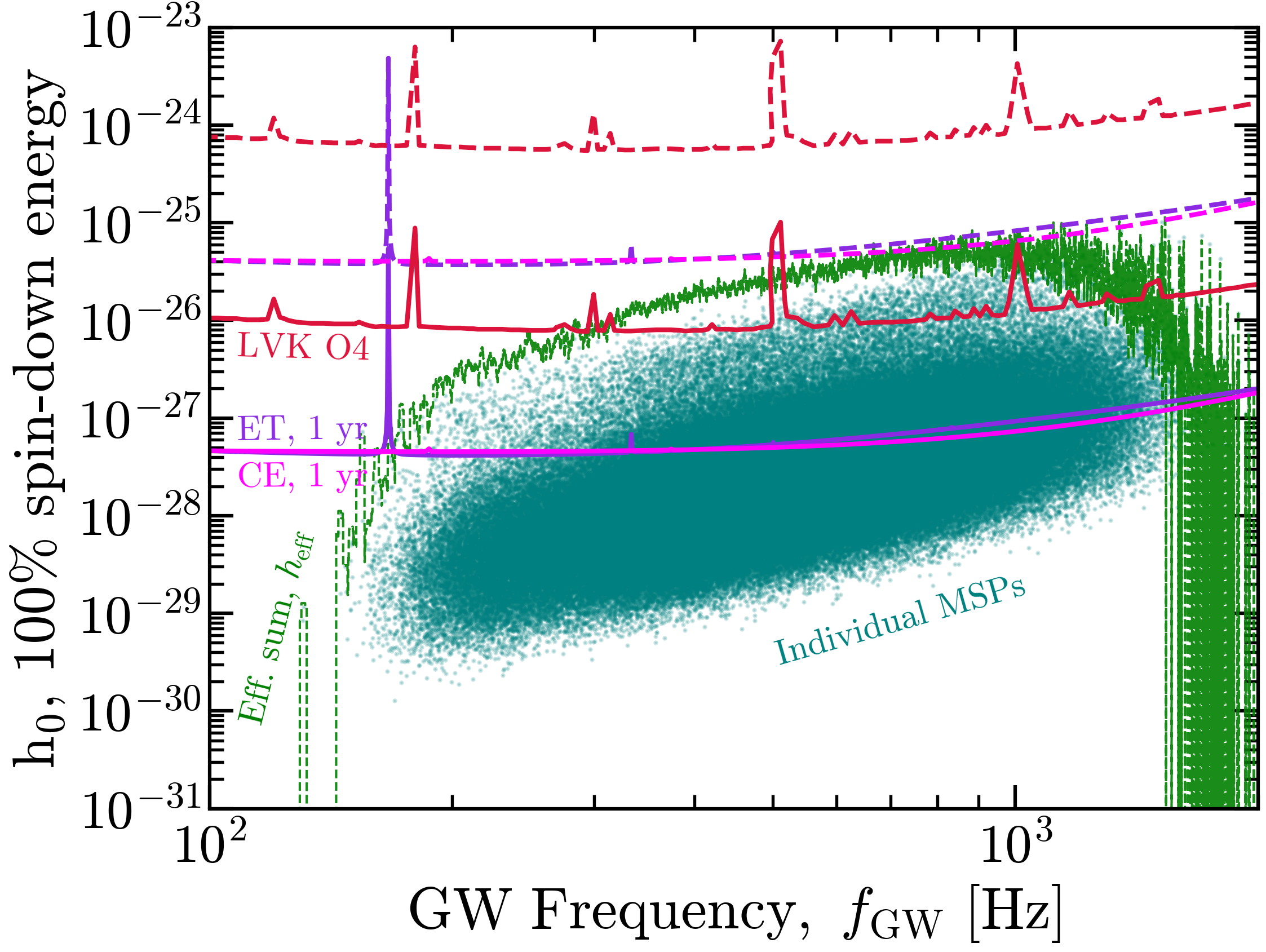}
\end{subfigure}
\begin{subfigure}[t]{\subfigwidth}
    \includegraphics[width=\rightpanelwidth]{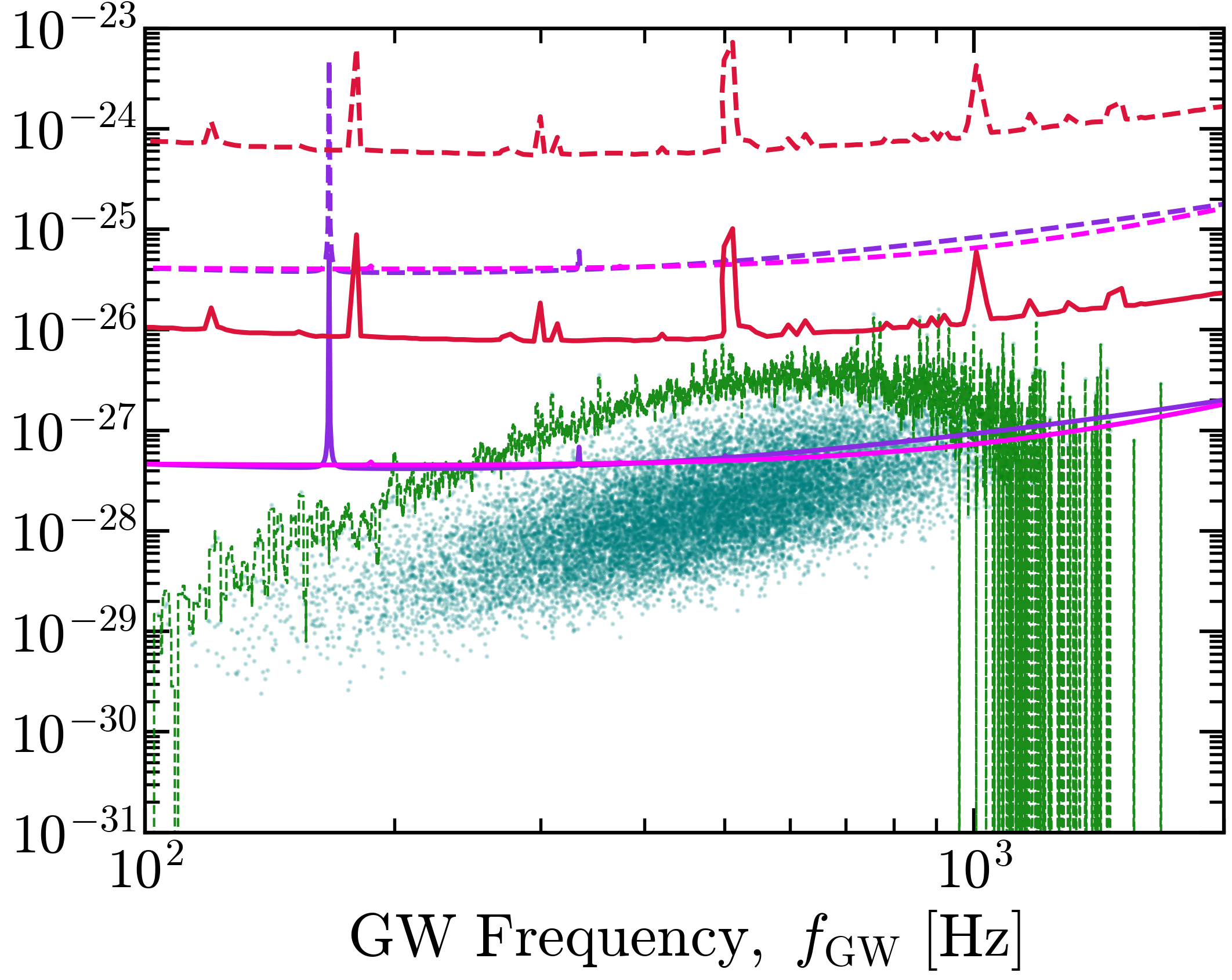}
\end{subfigure}
\\
\caption[OptionalShort]{\justifying
(Main results of the paper.) Our results for the intrinsic strain amplitude $h_0(f_{\rm GW})$ of bulge MSPs responsible for the GCE, shown for different ellipticity origins (vertical panels; Sec.~\ref{sec_ellipticity}) and MSP population models (horizontal panels; Sec.~\ref{sec_pop_paras}), along with the sensitivity for continuous wave searches from different GW detectors, as labeled.
\\
The {\bf {\color{teal}scattered points}} represent our predicted $h_0$ for the individual MSPs, to be compared with the \textbf{solid} sensitivity curves (coherent strategy; Eq.~\eqref{eq_h0min_coh}). 
The {\bf {\color{darkgreen} green dashed lines}} represent the total effective strain amplitude (Eq.~\eqref{eq_heff}) from the MSPs within a $2$-Hz bin, to be compared with the \textbf{dashed} sensitivity lines (incoherent strategy; Eq.~\eqref{eq_h0min_incoh}).
\\
{\bf Top row}: Ellipticity originating from magnetic-field-induced deformation (Sec.~\ref{sec_ellipticity_MF}).
{\bf Second row}: Ellipticity originating from crustal mountains from breaking strain (Sec.~\ref{sec_ellipticity_crustal}).
{\bf Third row}: Ellipticity assuming $\eta=1\%$ of spin-down energy converts to GW emission (Sec.~\ref{sec_ellipticity_Eloss}).
{\bf Bottom row}: \textit{Theoretical maximum}, assuming $\eta=100\%$ of spin-down energy converts to GW emission (Sec.~\ref{sec_ellipticity_Eloss}).
{\bf Left column:} MSP Population  \Rmnum{1} based on the ATNF pulsar catalog~\cite{ATNF_catalog}.
{\bf Right column:} MSP Population \Rmnum{2} based on Ploeg et al.~\cite{Ploeg_2020}.
}
\label{fig_h0}
\end{figure*}

Fig.~\ref{fig_h0} shows our results on the intrinsic strain amplitudes of the bulge MSPs responsible for the GCE, for different ellipticity origins (Sec.~\ref{sec_ellipticity}) and MSP population models (Sec.~\ref{sec_pop_paras}), along with the sensitivity of continuous GW searches from current and future ground-based GW detectors. 
The results in the left column, from our Pop.~\Rmnum{1} MSP population model based on the ATNF pulsar catalog~\cite{ATNF_catalog}, are generally higher than those in the right column, from our Pop.~\Rmnum{2} model based on Ploeg et al.~\cite{Ploeg_2020}, mainly because the former has a $N_{\rm MSP}$ about ten times larger. In addition, the distributions in $f_{\rm GW} = 2 f_{\rm rot}$ and ellipticity in the two population models also lead to a difference in the strain amplitude.

The relative placement of the scattered points with respect to the \textbf{solid} sensitivity curves~\footnote{The sensitivity curves for LVK O4 are adopted from Ref.~\cite{LIGOScientific:2025kei}. The PSD data files for CE (with the baseline 40 km detector) and ET (with 20 km arms)~\cite{Branchesi:2023mws,ET:2025xjr} are available for download at \url{https://dcc.cosmicexplorer.org/CE-T2000017/public} and \url{https://apps.et-gw.eu/tds/ql/?c=16492}, respectively.}
illustrates the prospects of detecting those individual MSPs in the coherent strategy, while in the incoherent strategy, detectability is assessed by comparing the green dashed line with the \textbf{dashed} sensitivity curves.
{\it Overall, for the {\bf coherent strategy}, the predicted $h_0$ values are generally below the reach of current detectors~\cite{LIGOScientific:2014pky,VIRGO:2014yos,KAGRA:2018plz} but can be detectable by next-generation detectors such as CE and ET~\cite{Evans:2023euw,Punturo:2010zz}.}
For the {\bf incoherent strategy}, only the theoretical maximum from Pop.~\Rmnum{1} marginally exceeds the sensitivity of CE and ET with a one-year observation time.

The scatter of the points is mainly due to the broadness of the ellipticity distribution (Fig.~\ref{fig_ellipticity_dist}), the Earth-MSP distance distribution from the morphological model (Fig.~\ref{fig_boxy_sphereical}), and the frequency distribution (Fig.~\ref{fig_para_dist}). 
This explains why the crustal mountains scenario (second row of Fig.~\ref{fig_h0}) exhibits the least scatter, since its ellipticity distribution is effectively a $\delta$ function.

It is worth noting that {\it the true detectability can be more optimistic than that indicated by our baseline results}, as the amplitudes are primarily controlled by (1) the total number of MSPs, $N_{\rm MSP}$, and (2) the physical quantities affecting MSP ellipticity in each scenario. 
First, the $h_0$ values of individual MSPs are primarily influenced by $N_{\rm MSP}$ through the spread of their distribution, such that a larger $N_{\rm MSP}$ results in more MSPs populating the high-$h_0$ tail in the coherent strategy. 
In the incoherent strategy, the effective summed strain $h_{\rm eff}$ scales directly with $N_{\rm MSP}$. 
Thus, while a larger $N_{\rm MSP}$ enhances detectability in both the coherent and incoherent strategies, it does so through different mechanisms.
For the second factor, we discuss each scenario separately in the following text. 

The top row of Fig.~\ref{fig_h0} shows the case in which ellipticity is sourced by strong internal magnetic fields misaligned with the rotation axis. 
All MSPs fall below the sensitivity of current detectors, but the most extreme sources in the high-$h_0$ tail are detectable with CE or ET for Pop.~\Rmnum{1}. 
Besides the influence of $N_{\rm MSP}$ discussed above, the overall normalization of $h_0$ and $h_{\rm eff}$ is proportional to $B_{\rm int}$, for which we adopt a conservative ratio of $150 B_{\rm surf}$ for a direct comparison with previous work (e.g., Refs.~\cite{Miller:2023qph, Bartel:2024jjj}). 
A higher value of $B_{\rm int}$ will enhance detectability, as $B_{\rm int}/B_{\rm surf}$ for MSPs is expected to be $10^2$--$10^4$~\cite{Mastrano:2011aa, Ciolfi:2013dta, Haskell:2015psa}.

The second row of Fig.~\ref{fig_h0} corresponds to the scenario in which the ellipticity is sustained by crustal deformations at the long-term residual level. We adopt a representative ellipticity of $10^{-9}$, which sets the overall normalization of both $h_0$ and $h_{\rm eff}$, which are also influenced by $N_{\rm MSP}$ discussed above. 
Because the ellipticity is fixed, the differences between the two population models arise solely from $N_{\rm MSP}$ and the distribution of $f_{\rm GW}$. The resulting $h_0$ values are generally smaller than those in the magnetic-field scenario, and their detectability is correspondingly more difficult, remaining beyond the sensitivity of current and next-generation detectors.

The third row of Fig.~\ref{fig_h0} shows results when a fraction $\eta$ of the spin-down energy is assumed to be emitted as GWs. 
Here, we take $\eta = 1\%$. Note that $\sqrt{\eta}$ sets the overall normalization of both $h_0$ and $h_{\rm eff}$; both are additionally shaped by the influence of $N_{\rm MSP}$ discussed above.
In this scenario, the MSPs in the high-$h_0$ tail can exceed the sensitivity of ET or CE.

Finally, the bottom row of Fig.~\ref{fig_h0} shows the theoretical maximum case, where $\eta = 100\%$, i.e., all spin-down energy is converted into GW emission. 
In this maximum scenario, both $h_0$ and $h_{\rm eff}$ are a factor of 10 larger than in the $\eta = 1\%$ case in the third row, representing the absolute upper bound on the GW emission from MSPs.
Remarkably, a fraction of the MSPs in the high-\(h_0\) tail exceed the LVK O4 sensitivity, implying that Advanced LIGO and Virgo at design sensitivity can potentially detect some individual sources~\cite{Capote:2024rmo,LIGOScientific:2025kei}. 
Conversely, a null detection in O4 would constrain such highly optimistic scenarios, placing meaningful upper limits on extreme ellipticities and restricting the parameter space for the MSP interpretation of the GCE.

The first two ellipticity scenarios in Fig.~\ref{fig_h0} are based on distinct physical mechanisms and may jointly contribute to the MSP ellipticity~\cite{Pagliaro:2023bvi}. The third and bottom rows provide model-independent parameterizations. The two population models further capture the uncertainties in MSP properties.
Therefore, the detectability shown in Fig.~\ref{fig_h0} should be interpreted as scenario-dependent forecasts over a broad range of astrophysical assumptions, rather than precise predictions. Nevertheless, the qualitative trends remain robust.

The solid (coherent) sensitivity curves in Fig.~\ref{fig_h0} represent a best-case bound. Doppler shifts from Earth's motion and binary orbits, combined with unknown MSP positions and limited computational resources, prevent full phase-coherent correction over a year-long observation. Conversely, the dashed (incoherent) curves serve as a worst-case bound, as phase information is lost between segments. The true detectability lies between these two limits.

A semicoherent strategy (not shown in the figure, for simplicity) provides a more realistic detection scheme targeting the GC~\cite{LIGOScientific:2013wcb,Piccinni:2019zub,KAGRA:2022osp}. By restricting the search to a single spatial bin centered on the GC, the coherent integration time per segment could realistically reach $T_{\rm stack} \sim 10^{4}$ s. Based on the established sensitivity scaling~\cite{KAGRA:2022osp,KAGRA:2022dwb,LIGOScientific:2026plm}, a one-year semicoherent search would degrade the baseline coherent strain sensitivity by a factor of $\sim 7$, while still being about $\sim 12$ times more sensitive than the incoherent strategy. Improved algorithms and greater computational resources may allow the semicoherent strategy to move closer to the coherent limit.

Our results are all based on the boxy bulge model. In Appendix~\ref{sec_app_morpho_dep}, we show that they are insensitive to the assumed bulge morphology.

Moreover, our results are based on the ROI of $|\ell| < 20^\circ$ and $2^\circ < |b| < 20^\circ$. However, there are likely many MSPs within $|\ell| < 20^\circ$ and $|b| < 2^\circ$. This Galactic plane region causes significant contamination in gamma-ray observations, but this may not be the case for GW observations. 
In Appendix~\ref{sec_app_incl_Blt2}, we show that including this region increases the total number of MSPs by 70\%, thereby leading to only a modest increase in the GW signal and its detectability.


\subsection{Comparison with previous work}
\label{sec_results_compare}
Our detectability is a few orders of magnitude lower compared with Ref.~\cite{Miller:2023qph}, 
the discrepancy primarily arising from the inclusion of PSR~J0537-6910 within their ATNF-catalog-driven population sampling~\cite{Miller_Zhao}.

PSR~J0537-6910 has quite different properties from MSPs. Although its period is 16.2~ms, only slightly larger than that of typical MSPs (1--10~ms), its period derivative is $\dot{P}=5.2\times 10^{-14}$, many orders of magnitude larger than those of typical MSPs ($10^{-21}$--$10^{-19}$). 
As a result, the spin-down power is $\dot{E}\simeq 5\times 10^{38}$~erg/s, enormously higher than typical MSPs ($10^{33}$--$10^{35}$~erg/s). The corresponding spin-down age is only $\simeq 4.9\times 10^{3}$~yr, and the derived surface magnetic strength is $\simeq 9.3\times 10^{11}$~Gs, both consistent with a young pulsar rather than an MSP ($\sim 1$~Gyr and $\sim 10^8$--$10^9$~Gs).  
Consequently, the derived ellipticity of $\simeq 10^{-6}$ for PSR~J0537-6910 forms the extreme high-end tail of the distribution in Ref.~\cite{Miller:2023qph} (see their Figs.~4 and 5), fundamentally dominating the expected signal amplitude.
We do not include PSR~J0537-6910 because its above properties indicate that it is a young pulsar rather than an MSP.
Moreover, PSR~J0537-6910 is located in the Large Magellanic Cloud (LMC) rather than in the Milky Way~\footnote{Despite not being a member of the MSP population involved in our Pop.~\Rmnum{1} samples, it remains a promising candidate for targeted GW searches~\cite{Fesik:2020tvn,LIGOScientific:2021yby,LIGOScientific:2020lkw,LIGOScientific:2025kei}.}.

One might ask whether a population of young pulsars similar to PSR~J0537-6910 could reside in the bulge region ($|\ell|<20^\circ$ and $2^\circ < |b| <20^\circ$) and thereby account for the GCE. Our estimate indicates that the expected number of such objects is $\ll 1$, as follows. 
Nearly all the young pulsars originate from Galactic core-collapse supernovae, occurring at a rate of $\simeq 0.02\ \mathrm{yr^{-1}}$. 
For $|\ell|<20^\circ$ (corresponding to Galactocentric radius $< 3$~kpc), the fraction of pulsar density is $\sim 0.1$, obtained by integrating a normalized standard pulsar birth surface density profile, modeled as a shifted gamma function~\cite{Yusifov:2004fr, Lorimer:2006qs}, up to 3~kpc.
For $|b|>2^\circ$ (i.e., Galactocentric height $z>0.3$~kpc), the fraction is $\sim 0.4$, derived by integrating an exponential vertical distribution (normalized) $\propto e^{-|z|/H}$ with the young pulsars' scale height of $H = 330$ pc~\cite{Lorimer:2006qs}.
Among the several thousand pulsars listed in the ATNF pulsar catalog~\cite{ATNF_catalog}, PSR~J0537-6910 has the highest spin-down power and is the only young pulsar with a rotational frequency comparable to MSPs. Thus, the fraction ($f_{\rm fast}$) of such extreme young pulsars is very small, i.e., $f_{\rm fast} \sim \mathcal{O}(1/N_{\rm pulsar}) \ll 1$ as an order-of-magnitude estimate, where $N_{\rm pulsar}$ is the total number of cataloged pulsars.
Combining the above factors and adopting a time window of 5 kyr (the age of PSR~J0537-6910), the expected number of comparable young pulsars in the bulge region is $\sim {\rm 5 ~ kyr} \times 0.02\ \mathrm{yr^{-1}} \times 0.1 \times 0.4 \times f_{\rm fast} = 4 f_{\rm fast} \ll 1$.

As a comparison to the above, including one such object in the bulge MSP population models would imply the existence of $\sim 500$ analogous young pulsars in Pop.~\Rmnum{1} and $\sim 30$ in Pop.~\Rmnum{2}, obtained by scaling the total predicted MSP numbers (Sec.~\ref{sec_pop_NMSP}) by the sample size of 619 MSPs in our ATNF-based empirical population model. These values exceed the above estimate, $4 f_{\rm fast}$, by orders of magnitude. Furthermore, Ref.~\cite{List:2025qbx} showed that even if the GCE is due to point sources, the total number of MSPs should be $\sim 10^{5}$, much larger than any plausible young-pulsar population.

In addition, a pulsar with its immense spin-down power, of which J0537-6910 exhibits the maximum among detected pulsars, would generate conspicuous observational signatures, such as a bright pulsar wind nebula (PWN). Indeed, the PWN powered by PSR~J0537-6910 is so prominent that it has been detected by both \textit{Fermi}-LAT~\cite{Fermi-LAT:2015qoh} and H.E.S.S.~\cite{HESS:2012bib}, despite its location outside our galaxy in the LMC. While these are indirect, such prominent features would increase the probability of identifying this kind of object. Despite this, no other candidate with comparable properties has been observed, which strongly implies the intrinsic rarity of such pulsars. This further illustrates that including PSR~J0537-6910 is inconsistent with a realistic bulge population model.

In the energy-loss-fraction scenario, after removing PSR J0537-6910, our detectability is higher than that in Ref.~\cite{Miller:2023qph}, as shown by comparing the right panel of our Fig.~\ref{fig_ellipticity_dist} with their Fig.~5.
This is because Ref.~\cite{Miller:2023qph} used $\epsilon \propto \eta$~\cite{Miller_Zhao} instead of $\epsilon \propto \sqrt{\eta}$ in Eq.~\eqref{eq_epsilon_from_eta}, which comes from energy conservation.

\vspace{0.8cm}
Next, compared with Ref.~\cite{Bartel:2024jjj}, our detectability in the coherent strategy is much better.
A direct comparison of the sensitivity curves in their Figs.~10 and 11 with the square root of the detector PSD, $\sqrt{S_n(f_{\rm GW})}$, from their cited literature reveals a numerical match. 
This equivalence indicates that their analysis uses the detector PSD directly without incorporating the coherent observation-time integration required for continuous GW searches~\cite{Bartel_Profumo}, which is included in our calculation through Eq.~\eqref{eq_h0min_coh}.

Another important difference concerns the effective strain of the MSP population. In Ref.~\cite{Bartel:2024jjj}, the collective spectrum is obtained by summing Gaussian-blurred individual GW signals, effectively corresponding to a linear sum of the individual contributions. This overestimates the collective strain amplitude for uncorrelated MSP phases, whereas we use a quadrature sum given in Eq.~\eqref{eq_heff}.

Furthermore, while Ref.~\cite{Bartel:2024jjj} adopts the $B_{\rm surf}$ distribution from Ploeg et al.~\cite{Ploeg_2020} (used in our Pop.~\Rmnum{2}), it takes the $P$ distribution from the ATNF pulsar catalog~\cite{ATNF_catalog} (used in our Pop.~\Rmnum{1)} instead of that from Ref.~\cite{Ploeg_2020}. 
In contrast, our Pop.~\Rmnum{2} adopts all the parameter distributions from Ref.~\cite{Ploeg_2020} consistently.

\section{Conclusion}
\label{sec_concl}

The GCE has long been considered a leading candidate signal for particle DM annihilation. However, an unresolved population of MSPs in the Galactic bulge offers the other compelling astrophysical explanation. Distinguishing between these two interpretations is challenging in electromagnetic observations due to source confusion, pulse broadening, and extinction. Gravitational waves, by contrast, provide a dust- and confusion-free probe: a steadily rotating, nonaxisymmetric MSP emits a nearly monochromatic GW at about twice its rotational frequency, within the sensitive band of ground-based GW detectors, with amplitude mainly determined by its ellipticity.

In this work, we systematically assess the GW emission expected from a population of MSPs associated with the GCE and evaluate its detectability with current and future GW detectors.
Table~\ref{summary_table} provides an overview of the analytical framework and key parameters.
Accordingly, we consider three major scenarios for the origin of MSPs’ ellipticity: (1) magnetic-field-induced deformation, (2) long-lived crustal mountains from breaking strain, and (3) a model-independent parameterization in which a fraction $\eta$ of the spin-down power is emitted as GWs, for which we use $\eta=1\%$ and set a \textit{theoretical maximum} at $\eta = 100\%$, i.e., the spin-down limit.

For the population, we constructed (1) an Empirical Disk-Proxy Model (Pop.~\Rmnum{1}) based on MSPs in the ATNF pulsar catalog~\cite{ATNF_catalog}, and (2) an Evolutionary Model (Pop.~\Rmnum{2}) following Ploeg et al.~\cite{Ploeg_2020}, in which bulge MSP properties are evolved from universal birth distributions. 
We adopted a bulge-tracing boxy bulge morphology and normalized the total numbers to the GCE luminosity, yielding baseline populations of $N_{\rm MSP} \simeq 2.9 \times 10^5$ (Pop.~\Rmnum{1}) and $\simeq 1.62 \times 10^4$ (Pop.~\Rmnum{2}) within our ROI after geometric rescaling.

We evaluate the detectability of the GW emission from those MSPs under two search strategies that bracket realistic analyses. In the coherent strategy, we assume that the Doppler modulations due to Earth’s rotation and revolution are correctable (which requires significant computational power), and that the MSP is either isolated or, if in a binary system, has orbital parameters that are sufficiently well constrained to correct the orbital Doppler modulation. Thus, a single Fourier bin collects the signal and the sensitivity follows the standard $h_{0,{\rm min}} \propto (1/T_{\rm obs})^{1/2}$ scaling (Eq.~\eqref{eq_h0min_coh}). 
In the incoherent strategy, uncorrected or unknown Doppler modulation broadens the signal power over a frequency window set by the maximal Doppler shift, motivating ``stack-slide'' strategies with a sensitivity penalty $\propto (T_{\rm obs}/T_{\rm stack})^{1/4}$ (Eq.~\eqref{eq_h0min_incoh}) and an effective summed strain $h_{\rm eff}$ from all MSPs falling in the same frequency window (Eq.~\eqref{eq_heff}).

Our main result, on the GW strain amplitudes and their detectability, is shown in Fig.~\ref{fig_h0}.
Overall, in the coherent strategy, the predicted strain amplitudes, $h_0$, generally lie below the reach of current LVK sensitivities, but can be detectable by next-generation detectors such as CE and ET. 
In contrast, in the incoherent strategy, only the theoretical maximum from Pop.~\Rmnum{1} marginally exceeds the sensitivity of CE and ET.
In addition, it is worth noting that the true detectability can be more optimistic than suggested by our baseline forecasts, since the strain amplitudes are primarily governed by two factors: (1) the total number of bulge MSPs, $N_{\rm MSP}$, which controls the extent of the high-$h_0$ tail and the summed $h_{\rm eff}$'s normalization, and (2) the physical mechanisms that determine the ellipticity in each origin, for which we adopt conservative assumptions.
Finally, in the theoretical maximum case with $\eta = 100\%$, where all spin-down energy is converted into GW emission, the LVK O4 can potentially detect a fraction of MSPs. 
Conversely, a null detection in LVK O4 would constrain such optimistic scenarios, with implications for the maximum sustainable ellipticities of MSPs and the parameter space of the population.

Given these astrophysical uncertainties, particularly in the ellipticity and bulge MSP population models, the detectability conclusions presented in this work should be interpreted as forecasts across a wide parameter space rather than precise predictions. 
Nevertheless, the qualitative trends remain robust. 
Under the key assumptions outlined in Table~\ref{summary_table}, current GW detectors are unlikely to detect such signals except in extreme ellipticity scenarios, whereas next-generation detectors such as ET and CE may probe the high-$h_0$ tail.

Compared to previous work~\cite{Miller:2023qph, Bartel:2024jjj}, we conduct a systematic study of the bulge MSPs in the frequency range of ground-based GW detectors. For example,
(1) We consider major scenarios in which the MSP ellipticity originates;
(2) We outline both coherent and incoherent searching strategies while taking into account the Doppler effects;
(3) We consider both isolated MSPs and those in the binary systems.
Our detailed analysis yields detectability estimates for the bulge MSPs that are significantly more optimistic than those of Ref.~\cite{Bartel:2024jjj}, but substantially less optimistic than those of Ref.~\cite{Miller:2023qph}; see Sec.~\ref{sec_results_compare} for details.

Our work also outlines a clear path forward. Directed continuous GW searches toward the GC, optimized for the MSPs' GW frequencies and leveraging various detection methods to manage computational cost, will keep improving current limits. 
Theoretical and observational inputs are also needed for the MSPs' intrinsic and population properties, especially their ellipticity.

In summary, GW observations offer a unique and complementary window into the Galactic bulge MSP population, overcoming the severe challenges faced in electromagnetic searches. 
This provides a crucial way to test the MSP interpretation of the GCE, which is the alternative (astrophysical) explanation to the particle DM interpretation. 
Looking ahead, the next generation of GW interferometers may reveal a long-sought MSP population near the Galactic center.
Conversely, a null detection would place stringent constraints on a specific combination of the population's abundance, rotation and ellipticity distributions, binary properties and search efficiency, progressively narrowing the parameter space of the bulge MSP population.
More broadly, continued improvements in GW sensitivity will provide increasingly stringent tests of MSP hypothesis, with growing power to distinguish between the astrophysical and DM interpretations of the GCE.
Either outcome will carry profound implications for the origin of the GCE, the role of recycled neutron stars in shaping Galactic high-energy phenomena, and the broader quest to disentangle conventional astrophysical sources from potential dark-matter and other new-fundamental-physics signatures.


\section*{Acknowledgments}
We are grateful for helpful discussions with Kayla Bartel, John Beacom, Emanuele Berti, Ilias Cholis, Marc Kamionkowski, Gordan Krnjaic, Andrew Miller, Benjamin Owen, Luca Reali, Nick Rodd, Tracy Slatyer, Yue Zhao, and Haowen Zhong.
B.\,Z. is supported by Fermi Forward Discovery Group, LLC under Contract No. 89243024CSC000002 with the U.S. Department of Energy, Office of Science, Office of High Energy Physics. X.H. is supported by the National Key Research and Development Program of China (2022YFF0503304), the National Natural Science Foundation of China (No. 12322302), the Project for Young Scientists in Basic Research of Chinese Academy of Sciences (No. YSBR-061), and the Chinese Academy of Sciences.

\onecolumngrid

\newpage 

\appendix


\section{Doppler Shifts due to Binary Motion}
\label{sec_app_doppler}

In Sec.~\ref{sec_detect_doppler}, we discuss the Doppler shift in the GW frequency due to Earth's rotation, Earth's revolution, and orbital motion in binary systems. 
While the first two contributions are well understood, the binary orbital motion varies significantly across binary systems, leading to substantial differences in the maximum Doppler shift. 

We determine a maximum Doppler shift in frequency, which applies to most of the MSPs in the binary systems from the ATNF pulsar catalog, by calculating the $(\Delta f)^{\max }_{\mathrm{bin}}$ for all the confirmed MSPs in binaries. 
Following the analyses in Refs.~\cite{Leaci_2015, Leaci:2016oja, LIGOScientific:2019yhl}, the orbital velocity of an MSP in the binary system is given by
\begin{equation}
    v_{\rm bin} = \frac{2\pi a_p}{P_{\rm bin}(1-e)},
\label{eq_v_binary}
\end{equation}
where $a_p$ is the orbital semimajor axis, $P_{\rm bin}$ is the binary orbital period, and $e$ the orbital eccentricity. 
Accordingly, the maximum Doppler shift is
\begin{equation}
   (\Delta f)^{\max }_{\mathrm{bin}} \simeq 2 \frac{v_{\text{bin}}}{c} {f_0}, 
\end{equation}
which occurs once every half orbital period.

\begin{figure}
\includegraphics[width=0.46\textwidth]{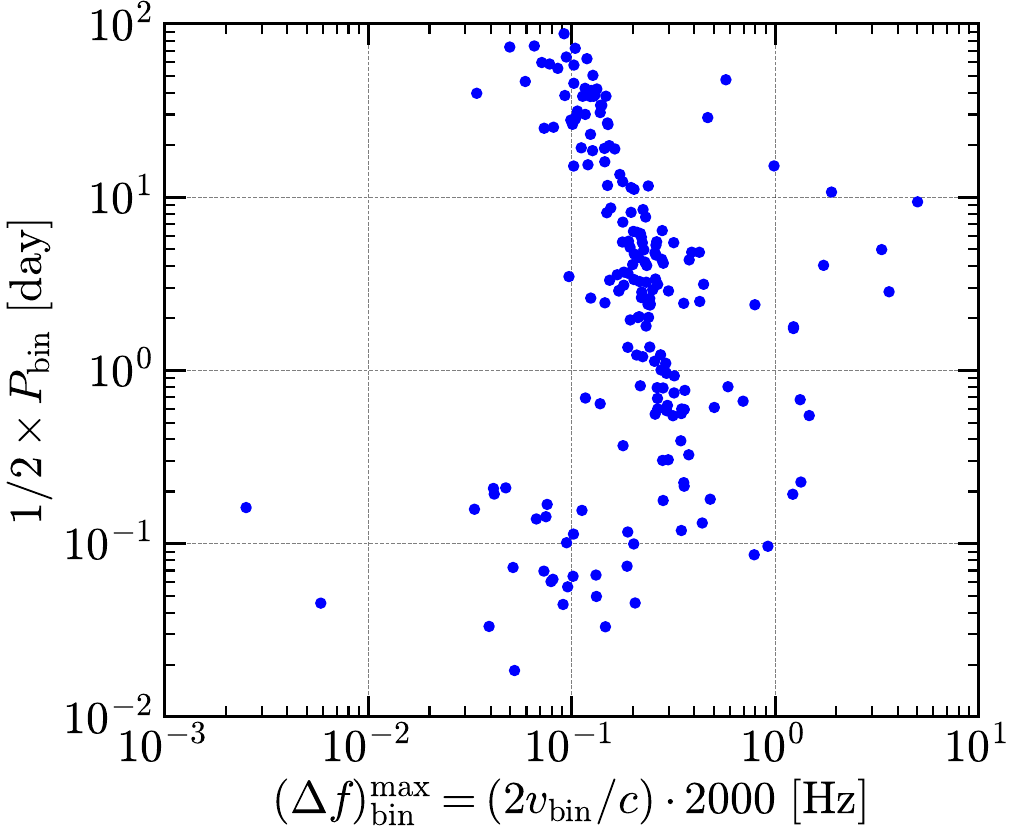}
\caption[OptionalShort]{\justifying
Distribution of maximum Doppler shifts, $(\Delta f)^{\max }_{\mathrm{bin}}$, versus half orbital period for the 199 MSPs in the binary systems from the ATNF pulsar catalog~\cite{ATNF_catalog}. 
Among the 619 MSPs used in the main text, 289 are confirmed to be in binary systems, of which 199 have measured orbital semimajor axis $a_p$, orbital period $P_{\rm bin}$, and eccentricity $e$, as required in Eq.~\eqref{eq_v_binary}.
}
\label{Fig_MSP_binary_Doppler}
\end{figure}

Fig.~\ref{Fig_MSP_binary_Doppler} shows the distribution of the maximum Doppler shift versus half orbital period for the 199 MSPs in binaries from the ATNF pulsar catalog~\cite{ATNF_catalog}.
For this sample, an integer value of $2$ Hz suffices to cover the maximum Doppler shift of about $95\%$ of the binaries, consistent with Eq.~\eqref{eq_Deltaf_max_binary}.
A small number of outliers exceed $2$ Hz; these generally correspond to systems with high orbital eccentricity $e$.

\section{GW Signals Insensitive to Assumed Bulge Morphological Models}
\label{sec_app_morpho_dep}

In this Appendix, we compare the GW results between different morphological distributions of the MSPs, which primarily affect the distances between the Earth and individual MSPs, which affects the GW strain, $h_0$ (Eq.~\eqref{eq_h0}). 
\textit{Our conclusion is that our GW results are nearly independent of the assumed morphological models of the bulge.
}

In the main text, we use the boxy bulge model only. Here, we adopt the other well-motivated model, the spherical bulge model. 
The spherical bulge follows the density profile~\cite{Binney:1996sv,Vanhollebeke_2009,Bartel:2024jjj}
\begin{equation}
    \rho_{\text{sph}} \propto \frac{\exp\left( - a^2 / a_m^2 \right)}{(1 + a/a_0)^{1.8}},
\label{spherical_bulge_distribution}
\end{equation}
where $a_m = 1.9$ kpc is the scale length, $a_0 = 0.1$ kpc is the inner truncation length, and $a \equiv \left(x^2 + y^2 / \eta^2 + z^2 / \zeta^2 \right)$~\cite{Binney:1996sv}. Here, ($x, y, z$) are Cartesian coordinates centered on the GC, and the axis ratios of the bulge are given by: $1 : \eta : \zeta = 1 : 0.68 : 0.31$~\cite{Vanhollebeke_2009}. 

\begin{figure}[htbp]
\centering
\begin{subfigure}[t]{0.49\textwidth}
\includegraphics[width=\linewidth]{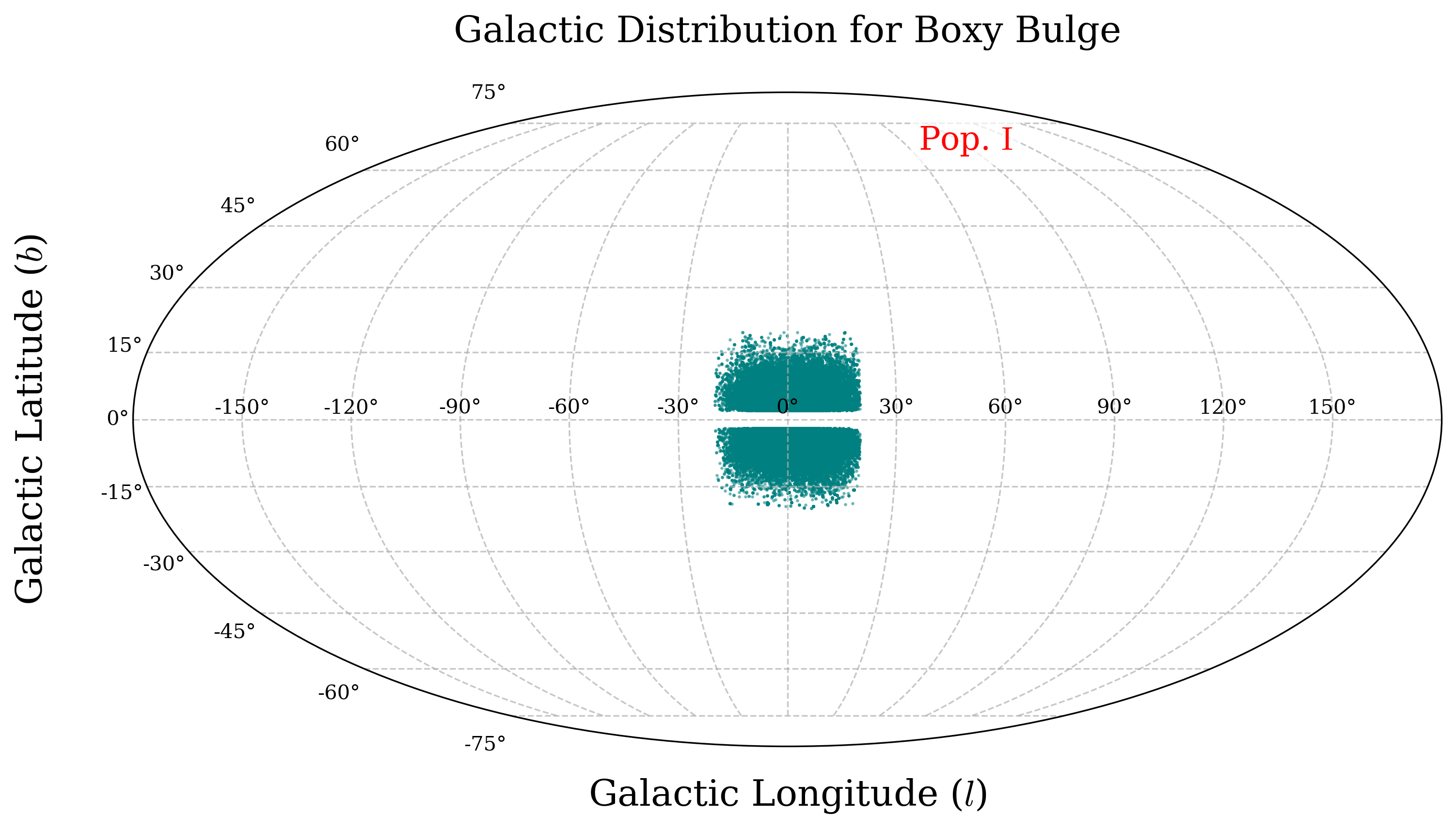}
\end{subfigure}
\begin{subfigure}[t]{0.49\textwidth}
\includegraphics[width=\linewidth]{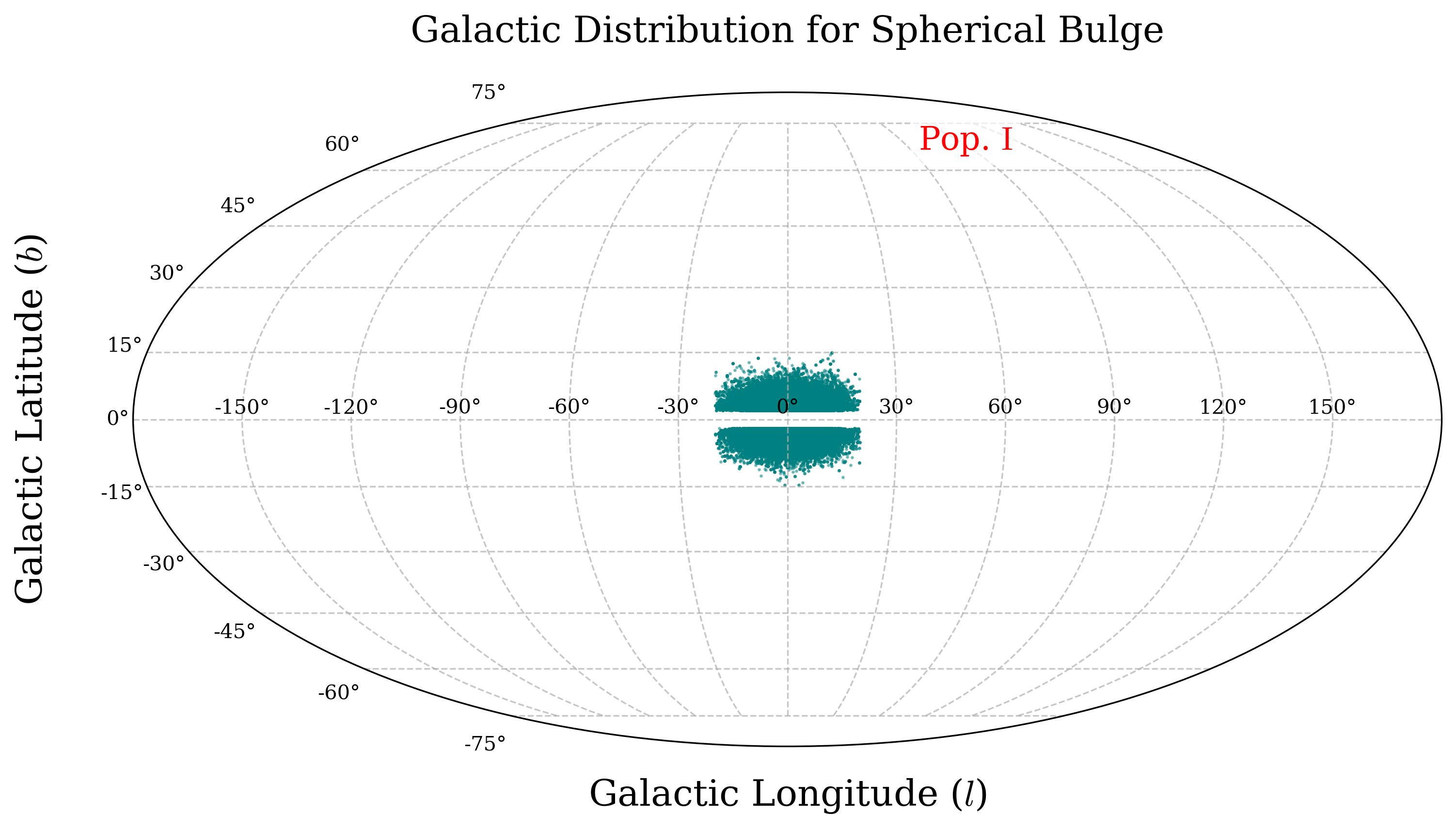}
\end{subfigure}
\caption[OptionalShort]{\justifying
Distribution of MSPs based on the boxy bulge model (\textbf{left}; used in the main text) and the spherical bulge model (\textbf{right}).
}
\label{fig_boxy_sphereical}
\end{figure}

Fig.~\ref{fig_boxy_sphereical} presents the two morphological distributions of the Galactic bulge: boxy (left panel) and spherical (right panel). The total number of MSPs is based on our Pop.~\Rmnum{1} discussed in the main text.

\begin{figure*}
\centering
\begin{subfigure}[t]{0.317\linewidth}
    \includegraphics[width=\linewidth]{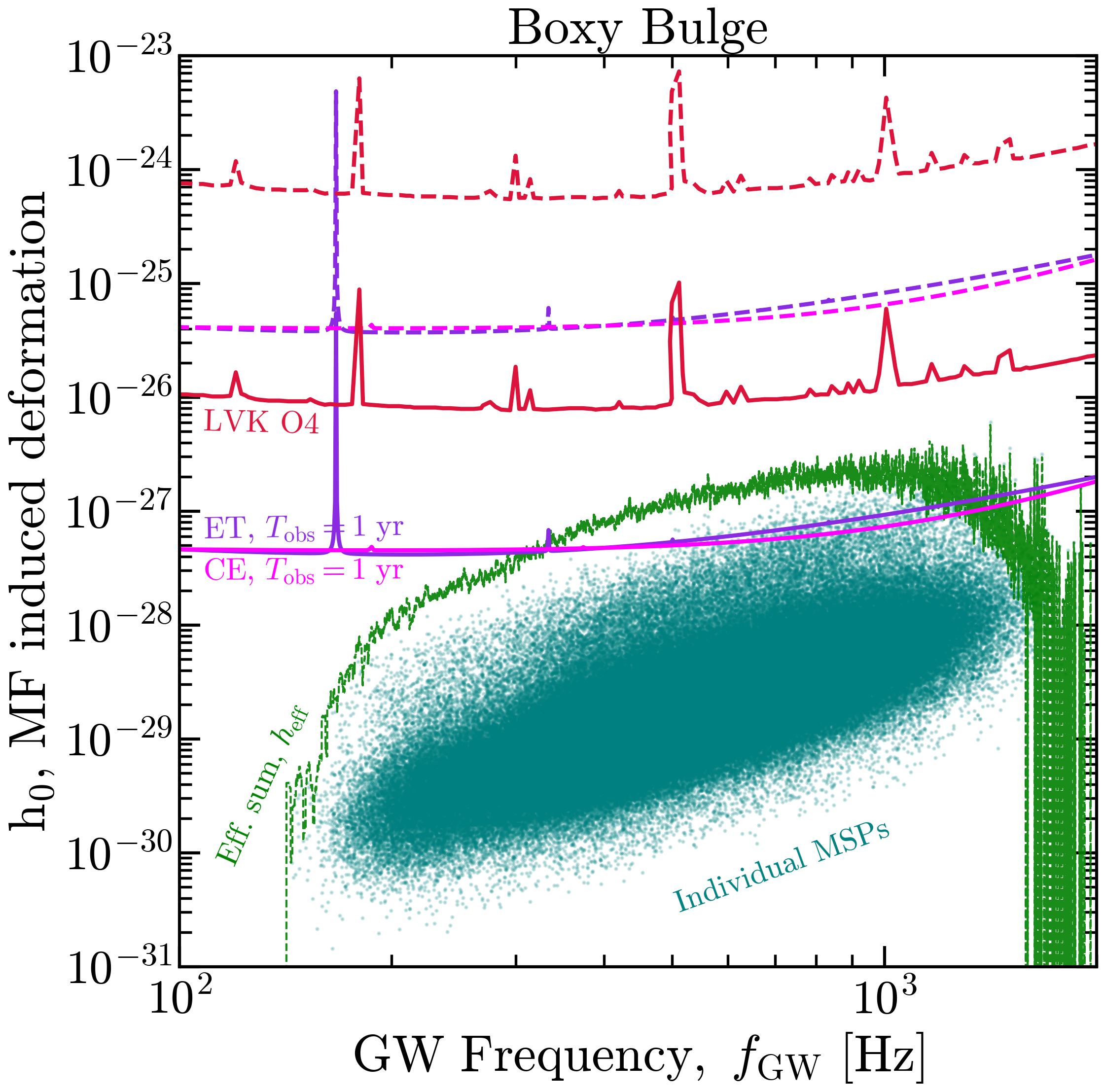}
\end{subfigure}
\begin{subfigure}[t]{0.30\linewidth}
    \includegraphics[width=\linewidth]{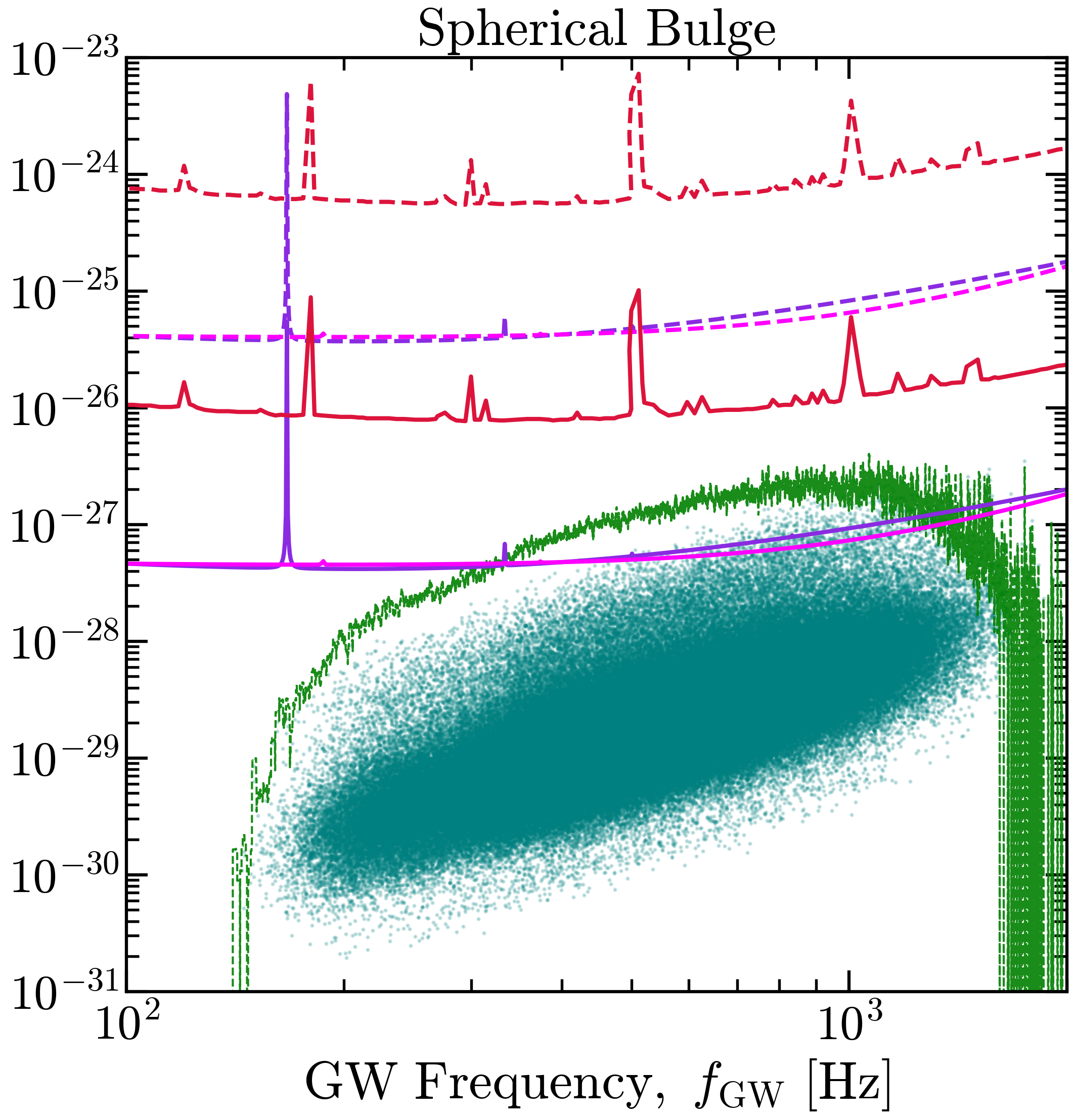}
\end{subfigure}
\begin{subfigure}[t]{0.30\linewidth}
    \includegraphics[width=\linewidth]{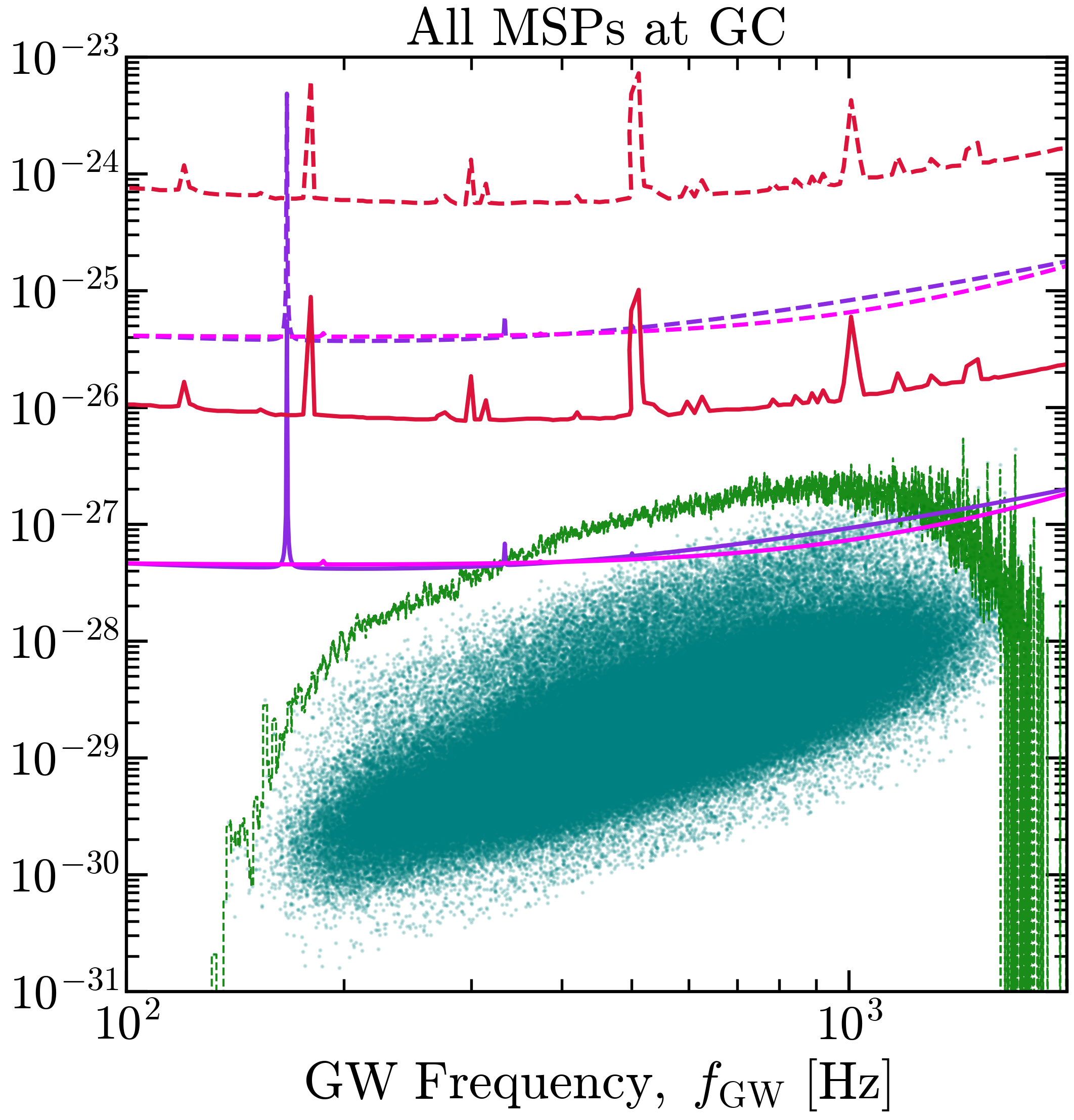}
\end{subfigure}
\caption[OptionalShort]{\justifying
Similar to Fig.~\ref{fig_h0} (main results of the paper) but comparing GW signals from the bulge MSPs between three different morphologies: \textbf{Left:} Boxy bulge, used in the main text and identical to the top-left panel of Fig.~\ref{fig_h0} (but plotted with a different figure size);
\textbf{Middle:} Spherical bulge;
\textbf{Right:} All MSPs placed at the GC.
Here, we show the ellipticity origin of magnetic-field-induced deformation, for simplicity, and the same conclusion applies to the energy-loss-fraction scenario.
For the crustal mountains scenario, the (unrealistic) all-MSPs-at-GC situation is different, but the boxy and spherical morphologies still give the same result.
See text for details.
}
\label{fig: h0_appendix_morphology}
\end{figure*}

Fig.~\ref{fig: h0_appendix_morphology} compares our GW results among different morphological distributions, including the boxy bulge (left panel) and the spherical bulge (middle panel) discussed above, plus an unrealistic extreme situation, i.e., all the MSPs are placed at the GC. 
Here, we show the scenario of magnetic-field-induced deformation.
Importantly, basically no difference is observed among the three morphological distributions. 
There are two main reasons. 
First, the one-dimensional distributions of Earth-MSP distances are more similar between the boxy and spherical morphologies than their corresponding two-dimensional spatial distributions.
Second, the ellipticity distribution (Fig.~\ref{fig_ellipticity_dist}, left panel) is much broader than the distance distribution.
For the energy-loss-fraction scenario, the same conclusion holds, since both reasons apply.
For the crustal mountains scenario, only the first reason applies, because the ellipticity distribution is a $\delta$ function (Fig.~\ref{fig_ellipticity_dist}, middle panel); still, the boxy and spherical bulges yield the same results.

\section{Results After Including the Galactic Plane}
\label{sec_app_incl_Blt2}

In our main text, we focus on the ROI of $\Omega_{\rm ROI}$ with $|\ell| < 20^{\circ}$, $2^{\circ} < |b| < 20^{\circ}$, in which the Galactic plane is masked. This is consistent with most GCE studies, as the Galactic plane introduces significant contamination in gamma-ray observations. 
However, this may not be the case for GW observations. Even in gamma-ray observations, some analyses of the GCE retain the Galactic plane~\cite{Macias_2019, Ploeg_2020}. 

Including the MSPs within $|b| < 2^\circ$ increases the total number of MSPs, $N_{\rm MSP}$, thereby enhancing their detectability. 
We calculate the total number of MSPs with $|b| < 2^\circ$ by scaling the $N_{\rm MSP}$ within $|\ell| < 20^{\circ}$ and $2^{\circ} < |b| < 20^{\circ}$, derived in Sec.~\ref{sec_pop_NMSP}, according to the boxy bulge morphology. 
The resulting $N_{\rm MSP}$ within $|\ell| < 20^{\circ}$ and $|b| < 20^{\circ}$ is $5 \times 10^5$ for Pop.~\Rmnum{1} and $27880$ for Pop.~\Rmnum{2}, about 70\% larger than the value obtained when masking $|b|<2^\circ$ in the main text.

\begin{figure*}
\centering
\begin{subfigure}[t]{\subfigwidth}
    \includegraphics[width=\linewidth]{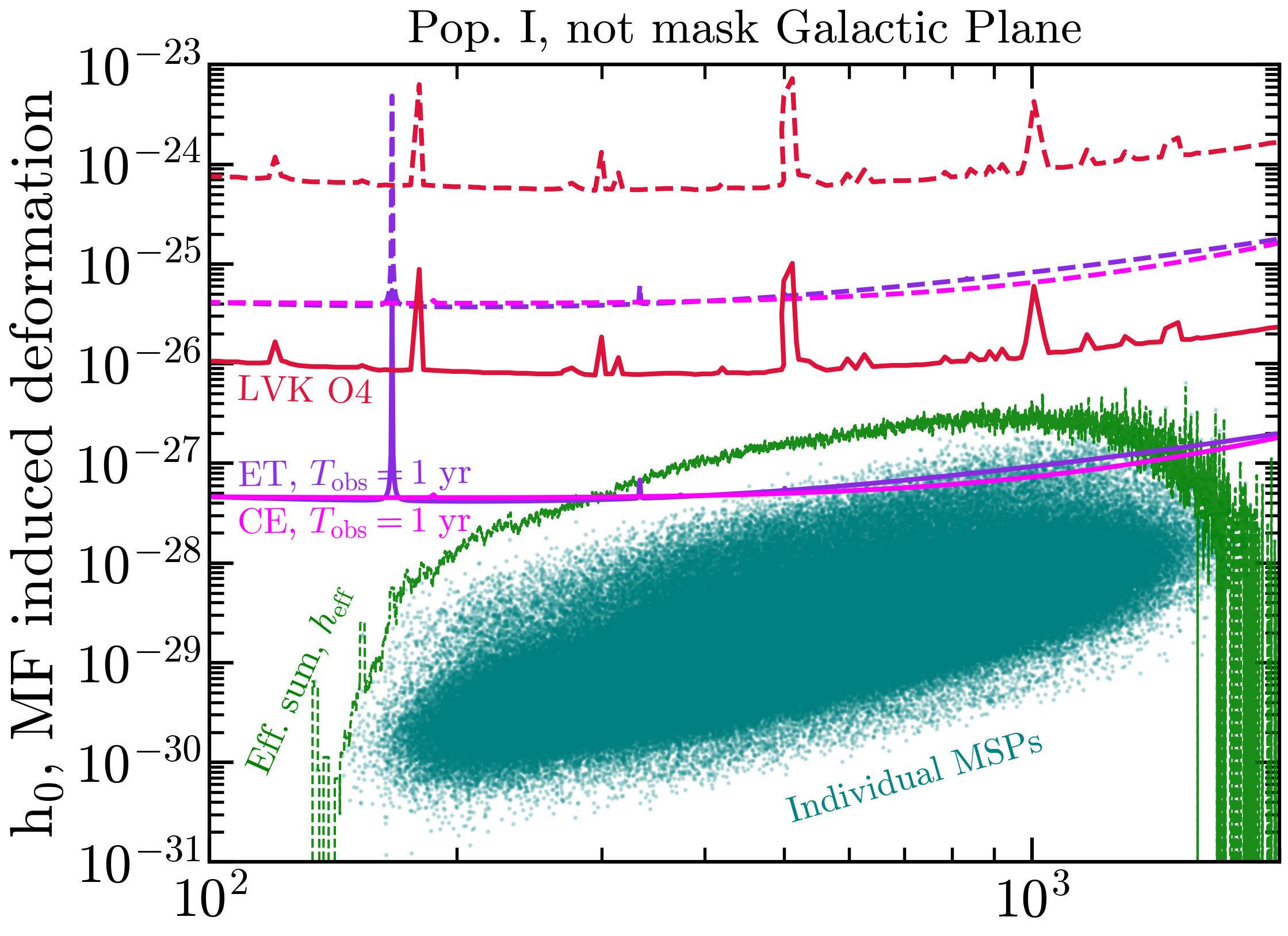}
\end{subfigure}
\begin{subfigure}[t]{\subfigwidth}
    \includegraphics[width=\rightpanelwidth]{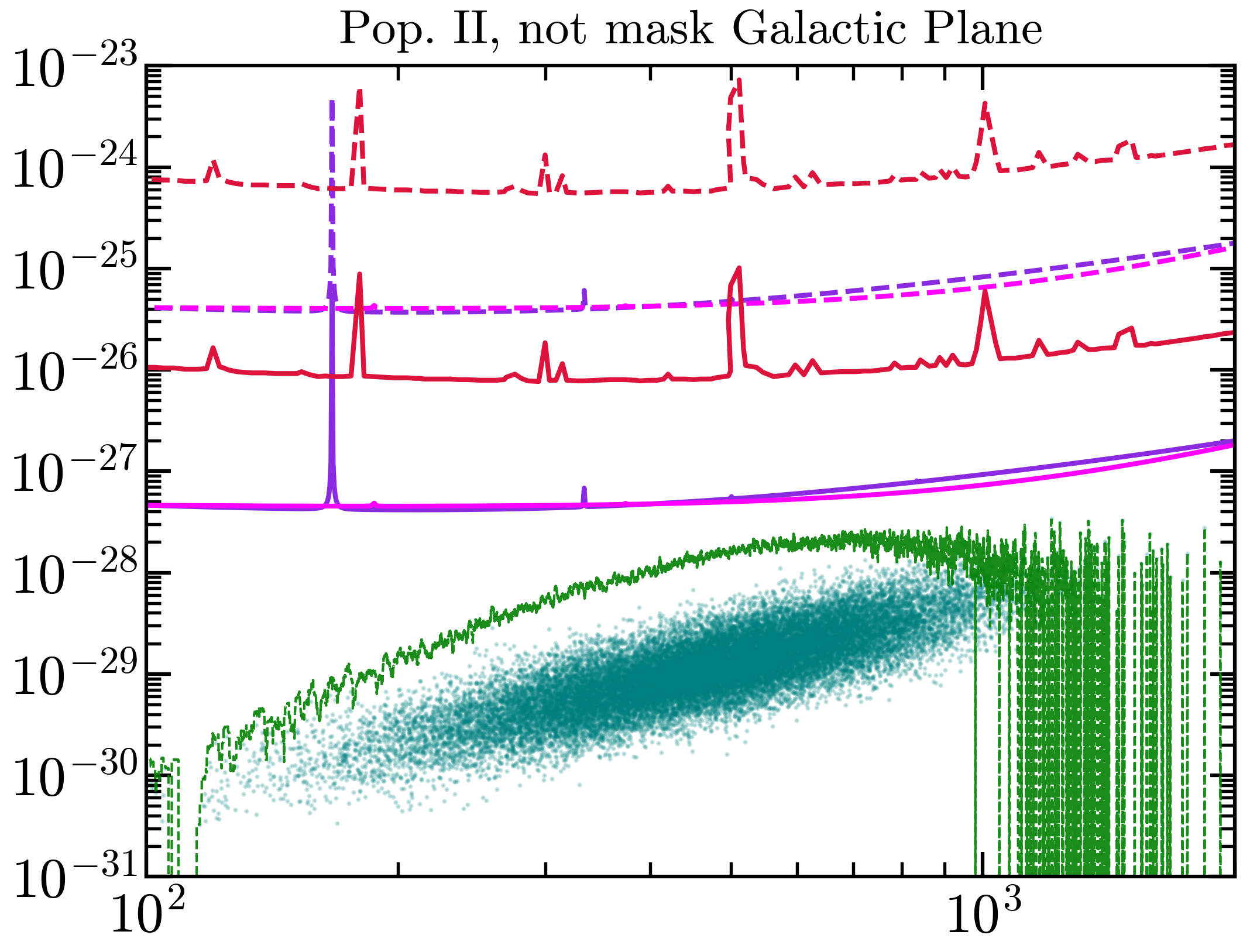}
\end{subfigure}
\\
\centering
\begin{subfigure}[t]{\subfigwidth}
    \includegraphics[width=\linewidth]{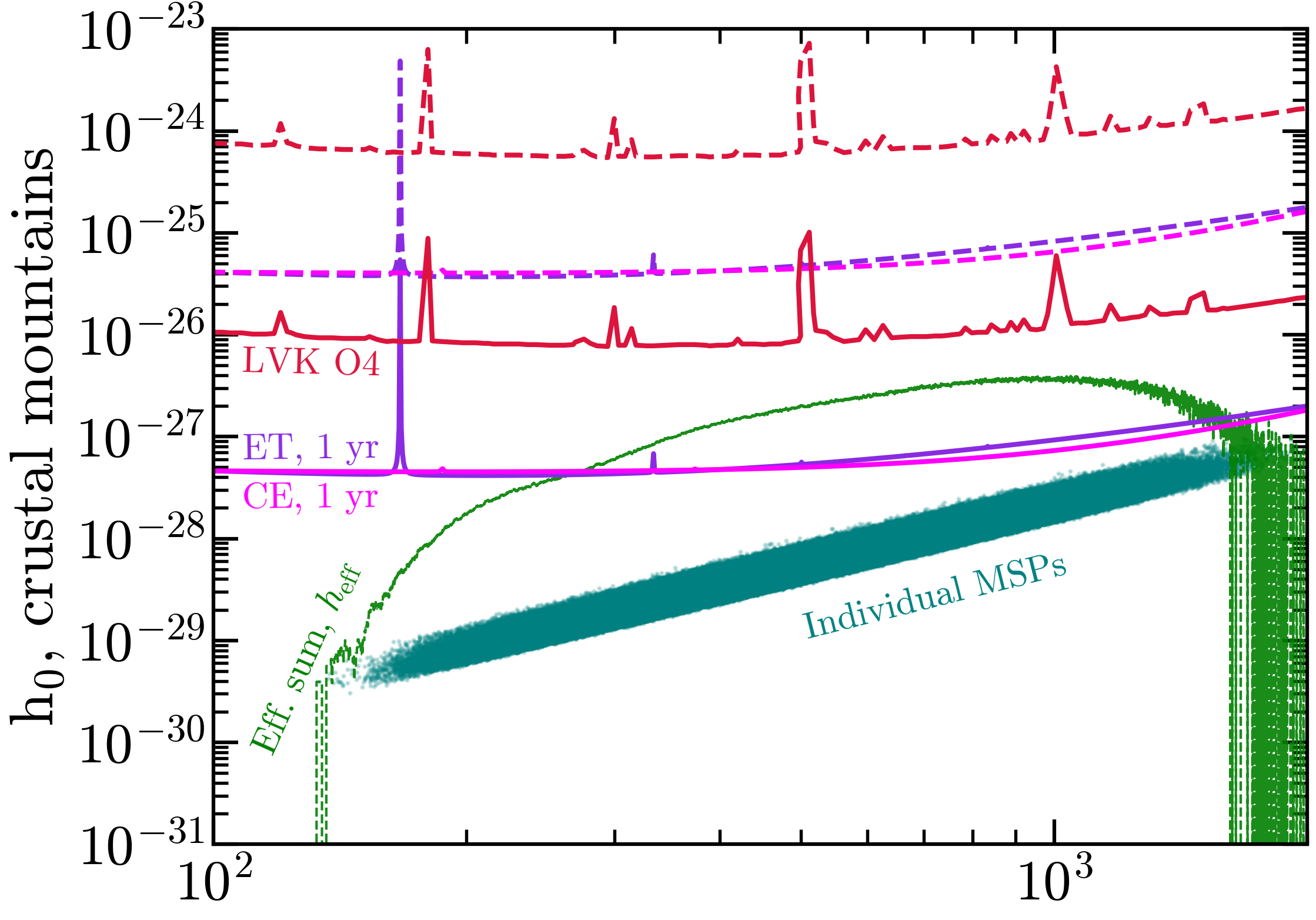}
\end{subfigure}
\begin{subfigure}[t]{\subfigwidth}
    \includegraphics[width=\rightpanelwidth]{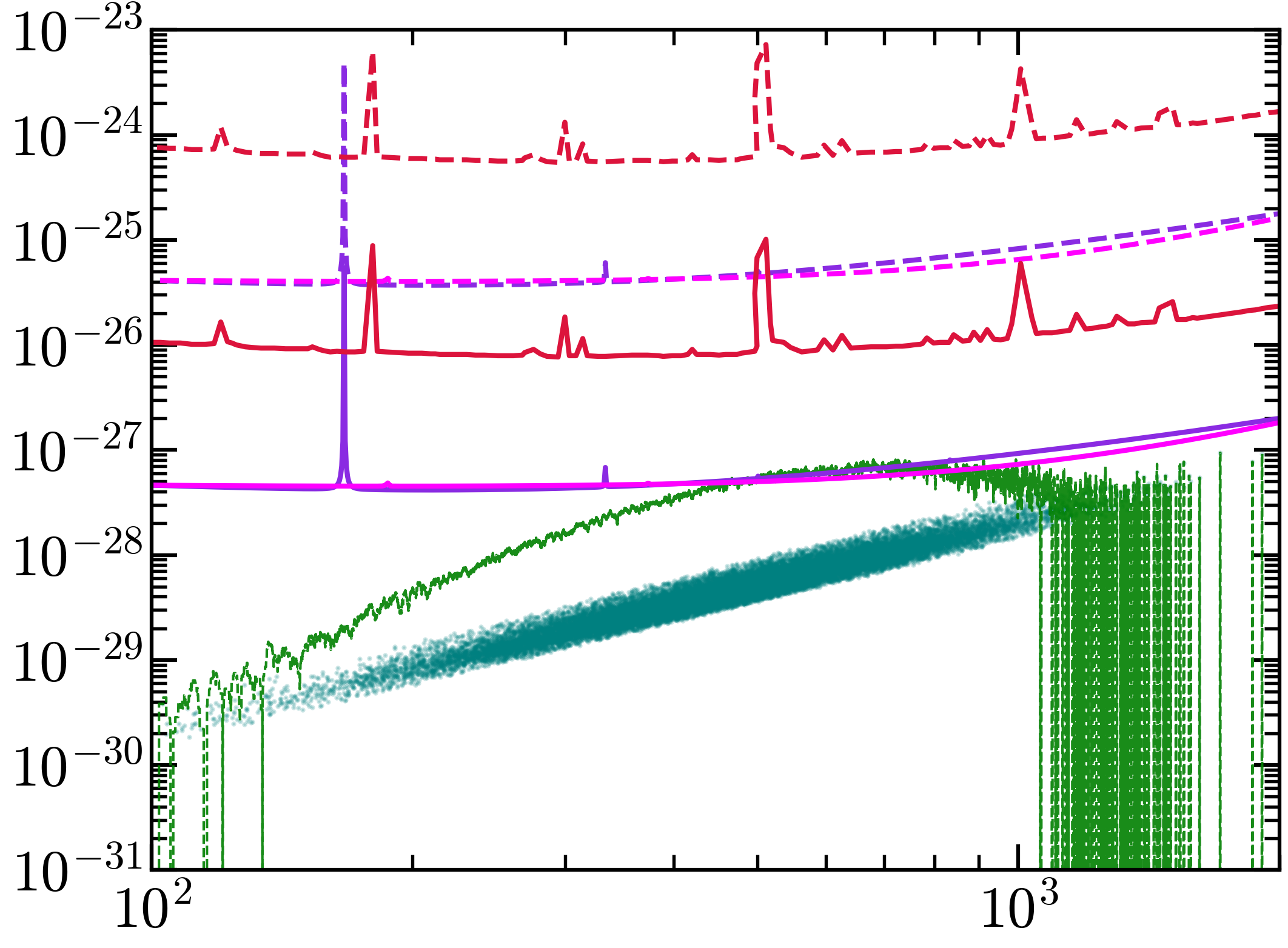}
\end{subfigure}
\\
\centering
\begin{subfigure}[t]{\subfigwidth}
    \includegraphics[width=\linewidth]{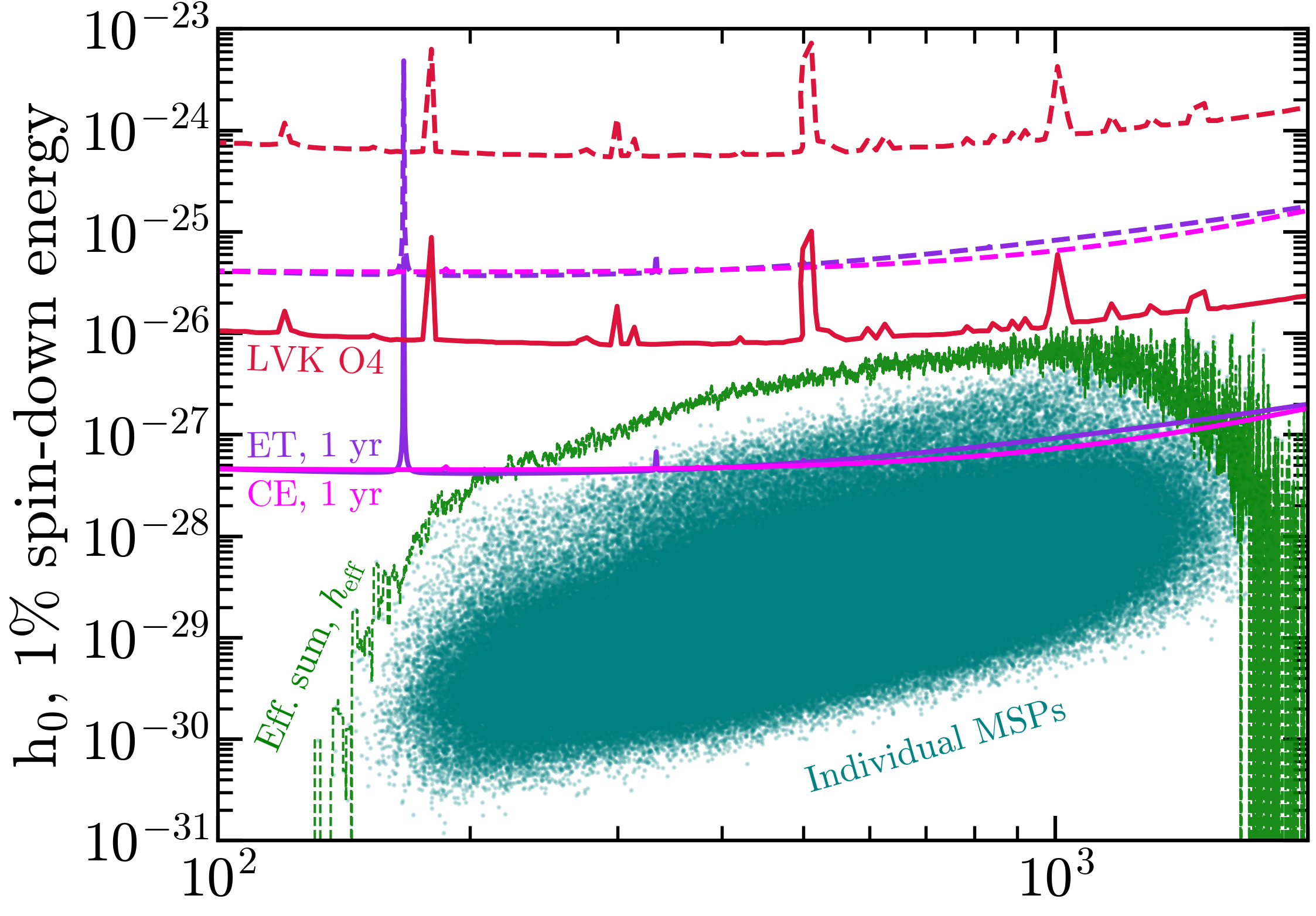}
\end{subfigure}
\begin{subfigure}[t]{\subfigwidth}
    \includegraphics[width=\rightpanelwidth]{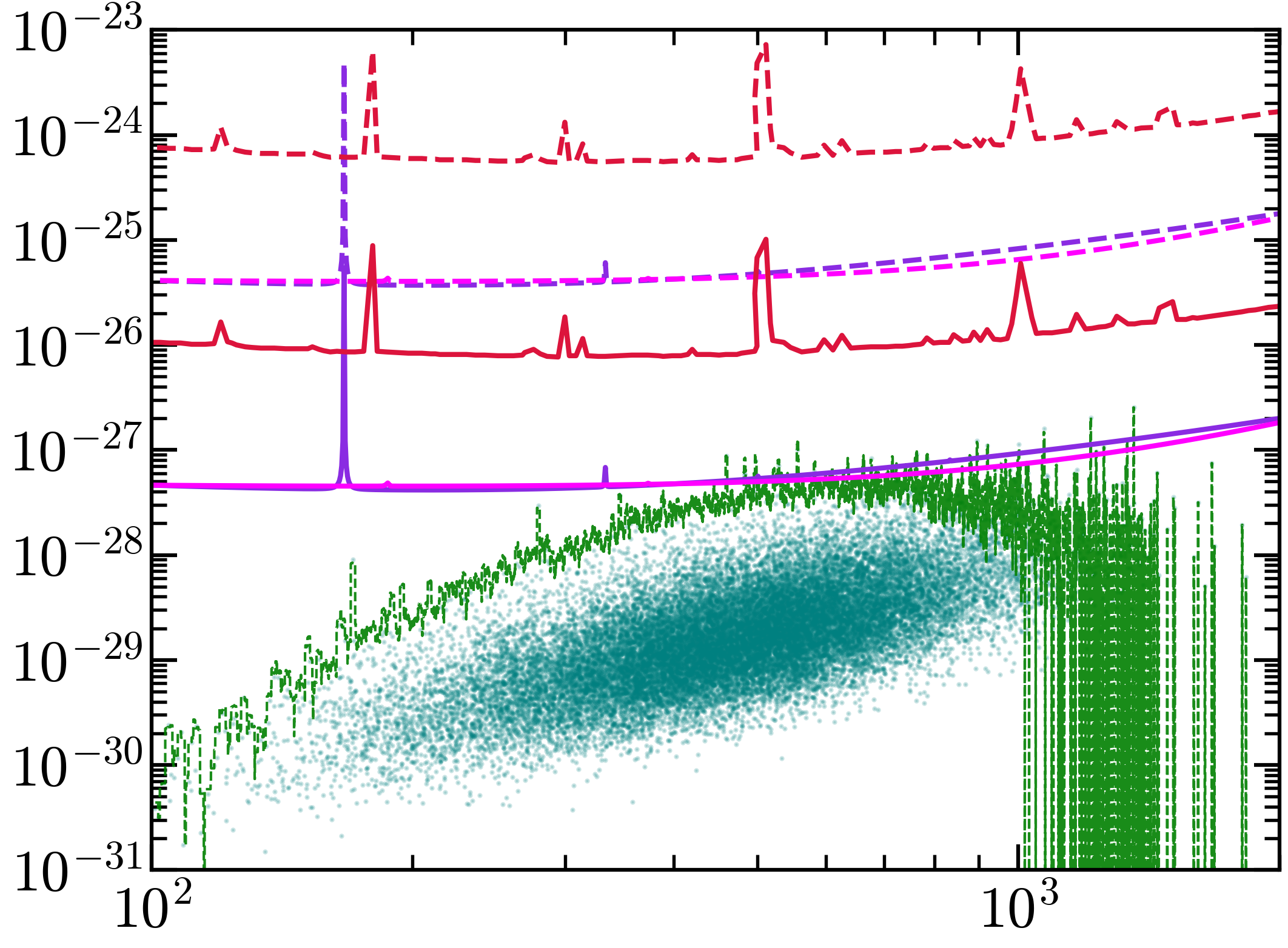}
\end{subfigure}
\\
\centering
\begin{subfigure}[t]{\subfigwidth}
    \includegraphics[width=\linewidth]{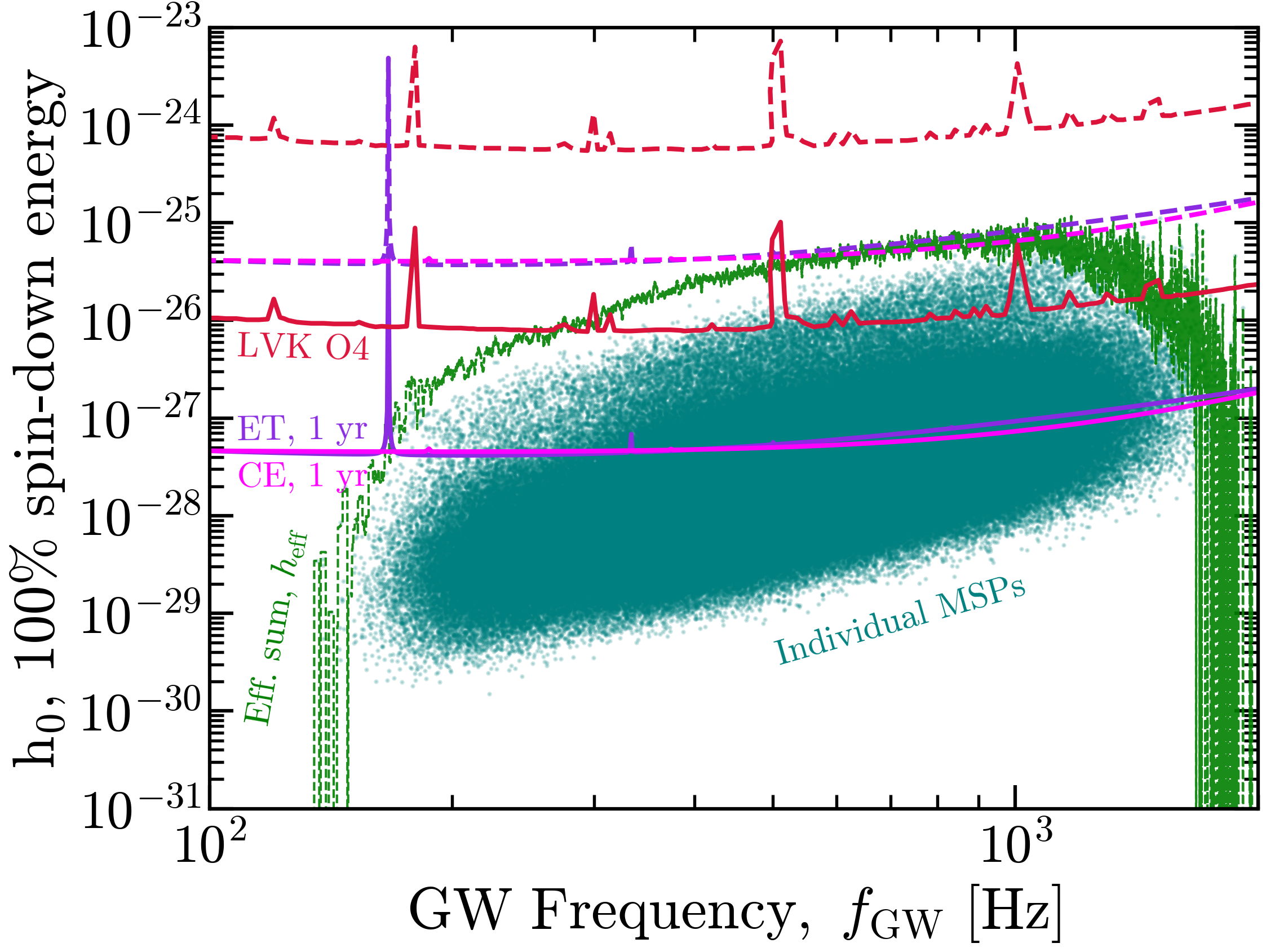}
\end{subfigure}
\begin{subfigure}[t]{\subfigwidth}
    \includegraphics[width=\rightpanelwidth]{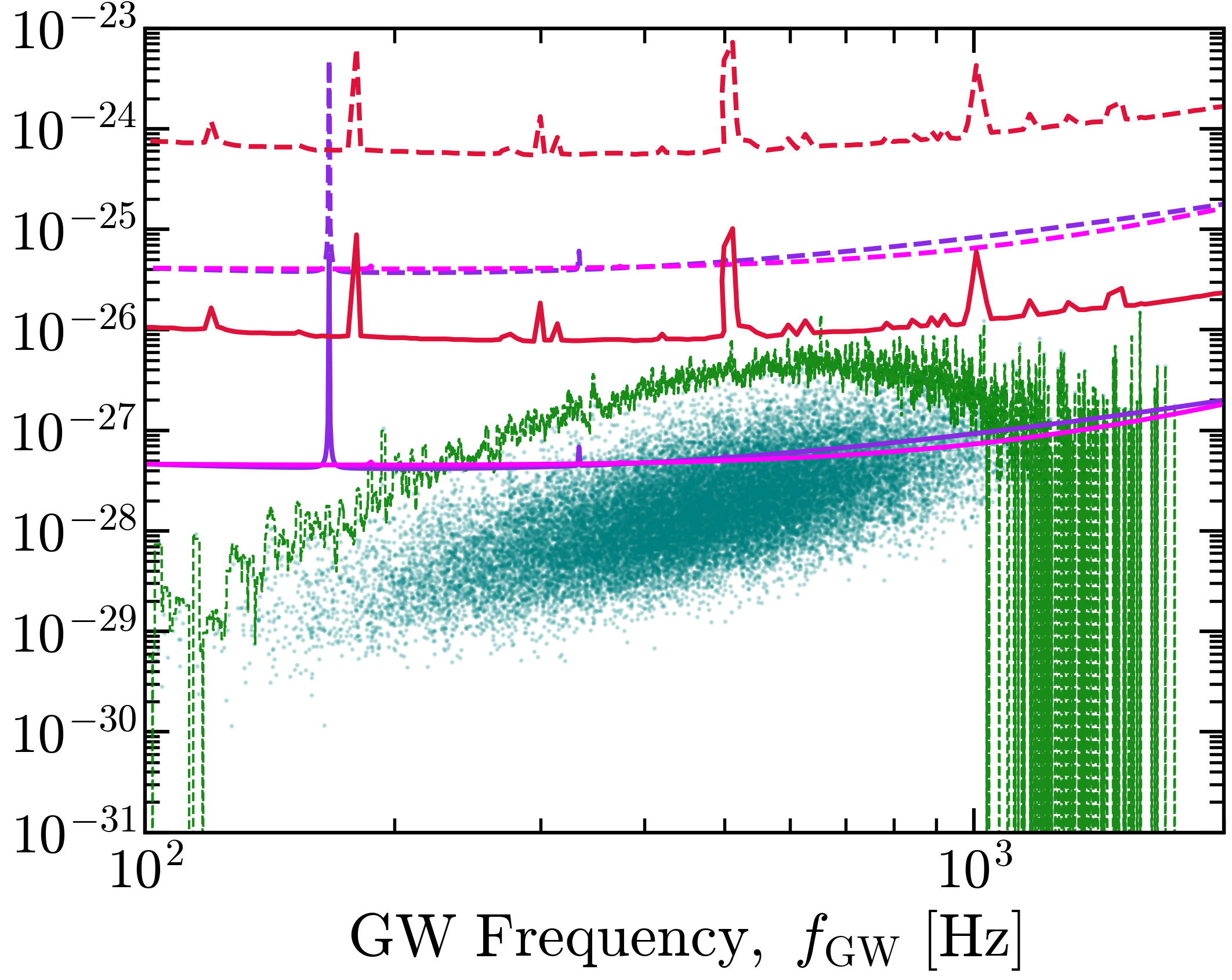}
\end{subfigure}
\\
\caption[OptionalShort]{\justifying
Same as Fig.~\ref{fig_h0} (main results of the paper), but without masking the Galactic plane ($|b|<2^\circ$), i.e., corresponding to the region $|\ell|<20^\circ$, $|b|<20^\circ$.
The Galactic plane causes significant contamination in gamma-ray detection, but may not pose a problem for GW detection.
Including $|b|<2^\circ$ and $|\ell|<20^\circ$ increases the total number of MSPs ($N_{\rm MSP}$) by 70\%, resulting in a modest improvement in GW detectability. See text for details.
}
\label{fig: h0_appendix_not_mask}
\end{figure*}

Fig.~\ref{fig: h0_appendix_not_mask} shows the GW results without masking $|b|<2^\circ$. 
Comparing with the main text results (Fig.~\ref{fig_h0}), we find that although $N_{\rm MSP}$ increases by 70\%, the improvement in GW detectability remains modest. 
This is because, for the coherent detection strategy, the increase in detectability arises from populating the high-$h_0$ tails of the distribution, and for the incoherent detection strategy, the enhancement is only $\sqrt{1.7} \simeq 1.3$.

\newpage
\twocolumngrid

\bibliography{References}

\end{document}